\documentclass[fleqn]{article}

\usepackage{amsmath,amssymb}
\usepackage{graphicx}
\usepackage{rotating}
\usepackage{geometry}
\usepackage{setspace}\setdisplayskipstretch{}
\usepackage{caption}
\usepackage{authblk}

\DeclareMathOperator{\real}{Re}
\DeclareMathOperator{\imag}{Im}
\DeclareMathOperator{\li2}{Li_{2}}

\renewenvironment{abstract}
{\small
 \begin{center}
 \bfseries \abstractname\vspace{-.5em}\vspace{0pt}
 \end{center}
 \list{}{%
  \setlength{\leftmargin}{3cm}%
  \setlength{\rightmargin}{\leftmargin}%
 }%
 \item\relax}
{\endlist}

\newcommand{\vpad}[2][4pt]{ \vtop{ \vbox{ \vspace*{#1} \hbox{$ #2 $}} \vspace*{#1} }}
\newcommand{\hgap}{\negthickspace}

\begin{document}

\title{A Spacing Estimator}
\author{Greg Kreider}
\affil{\small gkreider@primachvis.com}
\affil{\small Primordial Machine Vision Systems, Inc.}
\date{}

\maketitle

\begin{abstract}
\leftskip=5cm \rightskip=5cm
The distribution of the spacing, or the difference between consecutive order
statistics, is known only for uniform and exponential random variates.  We add
here logistic and Gumbel variates, and present an estimator for distributions
with a known inverse cumulative density function.  We show the estimator is
accurate to the limit of numerical simulations for points near the middle of
the order statistics, but degrades by up to 20\% in the tails.
\\
\leftskip=0pt \rightskip=0pt
\end{abstract}


\section{Spacings}
\label{sec:spacing}

Following Pyke's notation, the density $ D_{i} $of the spacing for a
distribution with density $ f(x) $ and cumulative density (c.d.f.) $ F(x) $ is
\cite[eq. (2.7)]{pyke65}
\begin{equation} \label{eq:fdi}
f_{D_{i}}(y) = \frac{n!}{(i-2)! (n-i)!}
  \int_{-\infty}^{\infty} \left\{ F(x) \right\}^{i-2} \left\{ 1 - F(x+y) \right\}^{n-i} f(x) f(x+y) dx
\end{equation}
This follows from the density of the order statistic, with the $i$'th value
factored out.  $ i $ is the index of the upper point, so that
$ 2 \leq i \leq n $ with $ n $ the number of points drawn.  The expected value
and variance follow normally from the moments of this density.
\begin{equation} \label{eq:edi}
E\bigl\{ D_{i} \bigr\} = \int_{0}^{\infty} y f_{D_{i}}(y) dy
\end{equation}
\begin{equation} \label{eq:vdi}
V\bigl\{ D_{i} \bigr\} = \int_{0}^{\infty} (y - E\bigl\{ D_{i} \bigr\})^2 f_{D_{i}}(y) dy
\end{equation}  

We have results for the spacing of the uniform and exponential distributions.
Details of the derivation are given in the Appendices; the following presents
just the results.

For uniform variates over an arbitrary range $ [a,b] $ the density function
and c.d.f. are
\begin{align} \label{eq:unif}
f(x) & = 1 / (b-a) & a \leq x \leq b \nonumber \\
F(x) & = (x-a) / (b-a) & a \leq x \leq b
\end{align}
with $ f(x) $ zero outside the range and $ F(x) $ zero below and one above.
These give a spacing density of
\begin{equation} \label{eq:fdi.unif}
f_{D_{i,unif}}(y) = \frac{n}{b-a} \left( \frac{b-y-a}{b-a} \right)^{n-1}
\end{equation}
The expected spacing is
\begin{equation} \label{eq:edi.unif}
E\bigl\{ D_{i,unif} \bigr\} = \frac{b-a}{n+1}
\end{equation}
and the variance of the spacing is
\begin{equation} \label{eq:vdi.unif}
V\bigl\{ D_{i,unif} \bigr\} =
  \frac{n}{(n+2)} \left( \frac{b-a}{n+1} \right)^{2}
\end{equation}
Taken over the unit range, these results match Pyke.  The variance is smaller
than the square of the expected value by $ n / (n+2) $.

For exponential variates with rate parameter $ \lambda $ the density function
and c.d.f. are
\begin{align} \label{eq:exp}
f(x) & = \lambda e^{-\lambda x} \nonumber \\
F(x) & = 1 - e^{-\lambda x}
\end{align}
The spacing density is
\begin{equation} \label{eq:fdi.exp}
f_{D_{i,exp}}(y) = \lambda (n-i+1) e^{-\lambda (n-i+1) y}
\end{equation}
This gives an expected spacing of
\begin{equation} \label{eq:edi.exp}
E\bigl\{ D_{i,exp} \bigr\} = \frac{1}{\lambda (n-i+1)}
\end{equation}
and variance
\begin{equation} \label{eq:vdi.exp}
V\bigl\{ D_{i,exp} \bigr\} = \frac{1}{\lambda^{2} (n-i+1)^{2}}
\end{equation}
Unlike uniform draws, the spacing for the exponential depends on the index
of the data point $ i $.  The exponential distribution is one-sided so the
spacing increases as $ i \rightarrow n $, or as we sample further into the
distribution's tail.  This variance is the square of the expected spacing.

We can also solve \eqref{eq:fdi}, \eqref{eq:edi}, and \eqref{eq:vdi} for
logistic variates.  This distribution takes two parameters, the location or
mean $ \mu $ and scale or standard deviation $ \sigma $.  The density function
and c.d.f are
\begin{align} \label{eq:logis}
f(x) & = \frac{e^{-z}}{\sigma (1+e^{-z})^{2}} \nonumber \\
F(x) & = \frac{1}{1 + e^{-z}} 
\end{align}
using $ z = (x - \mu) / \sigma $.  The density of the spacing follows from
\eqref{eq:fdi} using known integrals, giving
\begin{equation} \label{eq:fdi.logis}
f_{D_{i,logis}}(y) = \frac{1}{\sigma} e^{y/\sigma} \; \frac{(n-i+1) (i-1)}{n+1}
   \; {}_{2}F_{1}\left( i, n-i+2; n+2; 1-e^{y/\sigma} \right)
\end{equation}
The hypergeometric function converges only if $ |1 - e^{y/\sigma}| < 1 $,
which will not always be true.  An analytic continuation outside this region
works, and \eqref{eq:fdi.logis} matches a numeric integration of \eqref{eq:fdi}
except for $ y/\sigma \approx \ln 2 $.

We cannot substitute this directly into \eqref{eq:edi} and solve, but by
combining the density and expectation values into one equation and swapping
the integrals, so that the first integration over $ y $ includes the $ f(x+y) $
and $ F(x+y) $ factors from \eqref{eq:fdi}, we do find a closed form solution.
The expected value of the spacing is
\begin{equation} \label{eq:edi.logis}
E\bigl\{ D_{i,logis} \bigr\}
 = \frac{\sigma n!}{(i-2)! (n-i+1)!}
   \left\{ \left( \frac{1}{i-1} \right)^{2}
    - \sum_{k=1}^{n-i} \frac{(n-i-k)! (i-2)!}{(n-k)!} \right\}
\end{equation}
ignoring the summation if $ i = n $.

The variance of the spacing can be done, but the result is complicated.
Using an alternate form of \eqref{eq:vdi},
\begin{align} \label{eq:vdi2}
E\bigl\{ D^{2}_{i} \bigr\} & = \int_{0}^{\infty} y^{2} f_{D_{i}}(y) dy
 \nonumber \\
V\bigl\{ D_{i} \bigr\} & =
  E\bigl\{ D^{2}_{i} \bigr\} - \left[ E\bigl\{ D_{i} \bigr\} \right]^{2}
\end{align}
we can find a closed-form expression for the expected value of the square of
the spacing.
\begin{equation} \label{eq:edi2.logis}
E\bigl\{ D_{i,logis}^{2} \bigr\}
 = \frac{2 \sigma^{2} n!}{(i-2)! (n-i+1)!}
  \left\{ \begin{aligned}
    - \sum_{k=1}^{n-i} \frac{1}{k} \left( \frac{1}{-i+1} \right)^{2} \\
    + \sum_{k=2}^{n-i} \sum_{l=1}^{k-1} \frac{1}{k} \frac{(k-1-l)! (i-2)!}{(k-1-l+i)!} \\
    + \sum_{k=1}^{\infty} \frac{(-1)^{k}}{k^2} ( I_{out3B} - I_{out3A} ) \\
    - \frac{1}{-i+1} \frac{\pi^2}{12} \frac{1}{2^{i}} \\
    + \frac{1}{-i+1} \sum_{j=1}^{i-2} \frac{1}{i-1-j} \left[ \ln 2
     - \sum_{l=1}^{i-2-j} \frac{1}{i-1-j-l} \frac{1}{2^{i-1-j-l}} \right]
  \end{aligned} \right\} \\ \nonumber
\end{equation}
The intermediate result $ I_{out3B} - I_{out3A} $ depends on whether
$ k < i - 1 $, in which case
\begin{align} \label{eq:dilog1.logis}
I_{out3B} - I_{out3A}
 & = \frac{(-1)^{k} (i+k-2)!}{(k-1)!}
   \left[ \sum_{j=2}^{k} \frac{(-1)^{j}}{2^{i-1}} \frac{(j-2)!}{(i+j-2)!}
    + \frac{1}{(i-1)!} \left\{ \sum_{j=1}^{i-1} \frac{1}{j 2^{j}} - \ln 2
    \right\} \right] \nonumber \\
 & \qquad - \frac{k! (i-k-2)!}{2^{i-1}} \left\{
    \frac{2^{i-1} - 1}{(i-1)!} - \sum_{j=1}^{k} \frac{1}{j! (i-j-1)!} \right\}
\end{align}
or if it is not, then
\begin{align} \label{eq:dilog2.logis}
I_{out3B} - I_{out3A}
 & = \frac{(-1)^{k} (i+k-2)!}{(k-1)!}
  \left[ \sum_{j=2}^{k} \frac{(-1)^{j}}{2^{i-1}} \frac{(j-2)!}{(i+j-2)!}
   + \frac{1}{(i-1)!} \left\{ \sum_{j=1}^{i-1} \frac{1}{j 2^{j}} - \ln 2
   \right\} \right] \nonumber \\
 & \qquad - \frac{(-1)^{k} k!}{(k+1-i)!}
  \left[ \sum_{j=1}^{k} \frac{(-1)^{j}}{2^{i-1}} \frac{(j-i)!}{j!}
   + (-1)^{i} \frac{1}{(i-1)!} \left\{
   \sum_{j=1}^{i-1} \frac{1}{j 2^{j}} - \ln 2 \right\} \right]
\end{align}
Simulations suggest that the variance simplies to the square of the expected
spacing, as also happens for the exponential; however, we have not worked
through the series to see if this result is actually true.  The relationship
does not hold for the other distributions considered in this paper, including
the uniform as noted above and by further simulation the rest.

A solution for draws from a Gumbel distribution also follows from known
integrals.  The distribution takes two parameters, a location or mean $ \mu $
and scale $ \sigma $.  The density function and c.d.f. are
\begin{align} \label{eq:gumbel}
f(x) & = \frac{1}{\sigma} e^{-(z+e^{-z})} \nonumber \\
F(x) & = e^{-e^{-z} }
\end{align}
again using $ z = (x - \mu) / \sigma $.  The density of the spacing is
\begin{equation} \label{eq:fdi.gumbel}
f_{D_{i,gumb}}(y) = \frac{n!}{(i-2)! (n-i!)} \frac{e^{y/\sigma}}{\sigma}
 \sum_{k=0}^{n-i} \binom{n-i}{k}
   \frac{(-1)^{n-i+k}}{\left( e^{y/\sigma}(i - 1) + n - i + 1 - k \right)^{2}}
\end{equation}
and the expected value is
\begin{equation} \label{eq:edi.gumbel}
E\bigl\{ D_{i,gumb} \bigr\}
  = i \binom{n}{i} \sigma \left\{
  -\frac{1}{n-i+1} \ln(i-1)
  + \sum_{k=0}^{n-i} (-1)^{k} \binom{n-i}{k} \frac{1}{1+k} \ln(i+k)  \right\}
\end{equation}
The alternating series of logarithms translates to a single fraction with
odd and even $ k $ separated into the denominator and numerator (if $ n-i $
is even, otherwise numerator and denominator).  Individual factors are then
raised to a power given by their binomial coefficient, and to leading order
this does sum to close to one because over the range of $ k $ the even and
odd binomial coefficients each sum to the same value, balancing the polynomial
order of both.  An expression for the variance of the spacing is not possible.

The series for the logistic and Gumbel expected spacings are difficult to
evaluate, because the terms must nearly cancel to counteract the combinatorial
explosion.  The factorial factors in \eqref{eq:edi.logis} and
\eqref{eq:edi.gumbel} can reach $ O(10^{n/3}) $, but the final value is
$ O(1) $.  Performing the sum requires high precision libraries.  For the
logistic series this gives results that match numeric integration of the
base equation, \eqref{eq:edi}.  For the Gumbel series additional work is
needed to get the final form, as described in \ref{app:gumbel}.


\section{Spacing Estimator}
\label{sec:dq}

Other distributions do not have analytic expressions for their spacing:
\eqref{eq:fdi} does not reduce to known integrals for them, and without a
density function we cannot solve for the expected spacing or variance.
Some distributions, such as the normal, beta, $ \chi^{2} $, and $ t $,
do not have closed forms for their c.d.f.  Others, including Rayleigh,
Weibull, and Frechet, have higher powers of exponentials in their c.d.f.
so that the shift $ x + y $ does not separate.  But we can estimate the
expected spacing.

Drawing a random variable from a distribution can be done by passing a
uniform variate through the inverse c.d.f. $ F^{-1}(p) $; this is also
called the quantile function.  We know from \eqref{eq:edi.unif}, though,
that the $ n $ points are placed equally over the range [0,1].  If
$ p_{j} $ be these values, then the spacing on average will be the
difference of the mapped points.
\begin{equation} \label{eq:dq.diff}
\stackrel{\sim}{E} \bigl\{ D_{i} \bigr\} = F^{-1}(p_i) - F^{-1}(p_{i-1})
\end{equation}
Dividing by the uniform spacing we get a finite difference approximation
of the derivative.
\begin{align} \label{eq:dq.deriv}
\stackrel{\sim}{E} \bigl\{ D_{i} \bigr\} & =
 \frac{F^{-1}(p_i) - F^{-1}(p_{i-1})}{p_{i} - p_{i-1}} (p_{i} - p_{i-1})
 \nonumber \\
 & = \frac{dF^{-1}(p)}{dp} \Delta p
\end{align}
We call this the quantile estimator of the expected spacing.

For exponential variates the estimator is exact.  The inverse c.d.f. is
\begin{equation} \label{eq:Fexp}
F^{-1}(p) = - \frac{1}{\lambda} \ln(1-p)
\end{equation}
and differentiating gives
\begin{equation} \label{eq:dFexp}
\frac{dF^{-1}(p)}{dp} = \frac{1}{\lambda} \frac{1}{1-p}
\end{equation}
If
\begin{align} \label{eq:p}
p_{i} & = \frac{i-1}{n} \nonumber \\
\Delta p & = 1 / n
\end{align}
then
\begin{equation} \label{eq:dq.exp}
\stackrel{\sim}{E} \bigl\{ D_{i,exp} \bigr\}
  = \frac{1}{\lambda} \frac{1}{1-p} \Delta p
  = \frac{1}{\lambda (n - i + 1)}
\end{equation}
which is just \eqref{eq:edi.exp}.

For uniform variates $ \Delta p = 1/(n+1) $ in \eqref{eq:edi.unif}, which
differs from \eqref{eq:p}.  The explanation is usually made
(in \cite[]{pyke65}, for example) that the uniform distribution is bounded
on both sides, allowing a spacing for $ i = 1 $, but that the exponential is
bounded on only one, so there is one less interval.  However, we will see that
unbounded distributions still use \eqref{eq:p}.

For logistic variates the estimator is also exact.  The result is much simpler
than \eqref{eq:edi.logis}.  The inverse c.d.f. and its derivative are
\begin{align} \label{eq:Flogis}
F^{-1}(p) & = \mu + \sigma \ln \left( \frac{p}{1-p} \right) \nonumber \\
dF^{-1}(p)/dp & = \sigma \frac{1}{p (1-p)}
\end{align}
Then
\begin{equation} \label{eq:dq.logis}
\stackrel{\sim}{E}\bigl\{ D_{i,logis} \bigr\}
  = \frac{\sigma}{p (1-p)} \Delta p
  = \frac{\sigma}{\frac{i-1}{n} \left( 1 - \frac{i-1}{n} \right)} \frac{1}{n}
  = \frac{\sigma n}{(i-1) (n-i+1)}
\end{equation}
The series in \eqref{eq:edi.logis} simplifies by first putting everything on
a common denominator, which eliminates the $ n! $ and $ (i-1)! $ factors and
introduces products, and then combining one factor from the leading
(non-series) term with the last in the series, which builds a factorial that
ends in $ (n-i)! $.  The final cancellation gives \eqref{eq:dq.logis}.
Details are in \ref{app:match}.

The estimator cannot be exact for Gumbel variates.  The inverse c.d.f. and
its derivative are
\begin{align} \label{eq:Fgumbel}
F^{-1}(p) & = \mu - \sigma \ln\left( -\ln(p) \right) \nonumber \\
dF^{-1}(p)/dp & = \sigma \frac{1}{p} \frac{-1}{\ln(p)}
\end{align}
which gives as an estimator
\begin{equation} \label{eq:dq.gumbel}
\stackrel{\sim}{E}\left\{ D_{i,gumb} \right\} =
  \frac{-\sigma}{p \ln(p)} \Delta p =
  \frac{-\sigma}{(i-1) \ln\left( (i-1) / n \right)}
\end{equation}
The logarithm of $ i-1 $ here is in the denominator, not the numerator as in
\eqref{eq:edi.gumbel}, and there is no way to invert the value to potentially
match the two.

This is true for other distributions with an invertible c.d.f., from which we
can calculate the quantile estimator.  Table~\ref{tbl:edi} presents these
distributions.  They divide into three groups.  In one the estimator depends
on powers of $ p $; the group includes exponential, logistic, Laplace, and
Pareto variates.  The second group involves combinations of $ p $ and
$ \ln p $ and includes the Gumbel, Rayleigh, Weibull, and Frechet
distributions.  The third group has the exceptions, uniform variates with
constant spacing, and Cauchy, whose estimator uses the secant of $ p $.  In
the first group the exponential and Pareto distributions are one-sided so the
spacing increases only at large indices $ i $; the logistic and Laplacian
distributions are symmetric (as is the Cauchy).  The second group has
generally asymmetric distributions with longer tails and greater spacing,
usually at larger $ i $.  We use as default parameters zero mean or location
and unit scale or standard deviation, picking instead for the exponential
$ \lambda = 1 $; for the Pareto $ a = 4 $ and $ b = 1 $; for the Weibull
$ a = 5 $ and $ b = 1.5 $; and for the Frechet $ \lambda = 3 $, $ \mu = 0 $,
and $ \sigma = 1 $.  These values emphasize differences between the spacing.

\begin{sidewaystable}
\caption{\label{tbl:edi} Distributions with Invertible c.d.f}
\medskip
\centering
\begin{math}
\begin{array}{lccccc}
\hline
 & \text{pdf} & \text{cdf} & \text{inv cdf} &
   \text{derivative} & \text{expected spacing} \\
 & f(x) & F(x) & F^{-1}(p) & dF^{-1}(p)/dp &
   \stackrel{\sim}{E}\bigl\{ D_{i} \bigr\} \\
\hline
\text{Cauchy} &
 \vpad{ \frac{1}{\pi \sigma} \frac{1}{1 + z^{2} } } &
 \vpad{ \frac{1}{2} + \frac{1}{\pi} \tan^{-1} z } &
 \vpad{ \mu + \sigma \tan \pi \left( p - \frac{1}{2} \right) } &
 \vpad{ \pi \sigma \sec^{2} \pi \left( p - \frac{1}{2} \right) } &
 \vpad{ \frac{\pi \sigma}{n} \sec^{2} \pi \left( \frac{i-1}{n} - \frac{1}{2} \right) } \\
\text{exponential} &
 \vpad{ \lambda e^{-\lambda x} } &
 \vpad{ 1 - e^{-\lambda x} } &
 \vpad{ - \frac{1}{\lambda} \ln(1-p) } &
 \vpad{ \frac{1}{\lambda} \frac{1}{1-p} } &
 \vpad{ \frac{1}{\lambda (n-i+1)} } \\
\text{Frechet} &
 \vpad{ \frac{\lambda}{\sigma} z^{-1-\lambda} e^{-z^{-\lambda}} } &
 \vpad{ e^{-z^{-\lambda}} } &
 \vpad{ \mu + \sigma \left[ -\ln(p) \right]^{\frac{1}{\lambda}} } &
 \vpad{ \frac{\sigma}{\lambda} \frac{1}{p} \biggl[ \frac{-1}{\ln(p)} \biggr]^{\frac{\lambda+1}{\lambda}} } &
 \vpad{ \frac{\sigma}{\lambda} \frac{1}{i-1} \biggl[ \frac{-1}{\ln\left( (i-1) / n \right)} \biggr]^{\frac{\lambda+1}{\lambda}} } \\
\text{Gumbel} &
 \vpad{ \frac{1}{\sigma} e^{-\left( z + e^{-z} \right)} } &
  \vpad{ e^{-e^{-z}} } &
 \vpad{ \mu - \sigma \ln\left( -\ln(p) \right) } &
 \vpad{ \sigma \frac{1}{p} \frac{-1}{\ln(p)} } &
 \vpad{ \sigma \frac{1}{i-1} \frac{-1}{\ln \left( (i-1) / n \right)} } \\
\text{Laplace} &
 \vpad{ \frac{1}{2 \sigma} e^{-|z|} } &
 \vpad{ \begin{cases} \frac{1}{2} e^{z} & x \leq \mu \\
                      1 - \frac{1}{2} e^{-z} & x \geq \mu
        \end{cases} } &
 \vpad{ \begin{cases} \mu + \sigma \ln 2p & p \leq 1/2 \\
                      \mu - \sigma \ln 2(1-p) & p \geq 1/2
        \end{cases} } &
 \vpad{ \begin{cases} \frac{\sigma}{p} & p \leq 1/2 \\
                      \frac{\sigma}{1-p} & p \geq 1/2
        \end{cases} } &
 \vpad{ \begin{cases} \frac{\sigma}{i-1} & i \leq (n/2)+1 \\
                      \frac{\sigma}{n - i + 1} & i \geq (n/2)+1
        \end{cases} } \\
\text{logistic} &
 \vpad{ \frac{e^{-z}}{\sigma \left( 1 + e^{-z} \right)^{2}} } &
 \vpad{ \frac{1}{1 + e^{-z}} } &
 \vpad{ \mu + \sigma \ln \left( \frac{p}{1-p} \right) } &
 \vpad{ \sigma \frac{1}{p (1-p)} } &
 \vpad{ \sigma \frac{n}{(i-1) (n-i+1)} } \\
\text{Pareto} &
 \vpad{ \frac{a b^{a}}{x^{a+1}} } &
 \vpad{ 1 - \left( \frac{b}{x} \right)^{a} } &
 \vpad{ b (1-p)^{-\frac{1}{a}} } &
 \vpad{ \frac{b}{a} (1-p)^{-\frac{a+1}{a}} } &
 \vpad{ \frac{b}{a} n^{\frac{1}{a}} (n-i+1)^{-\frac{a+1}{a}} } \\
\text{Rayleigh} &
 \vpad{ \frac{x}{\sigma^{2}} e^{-\frac{1}{2} \left( \frac{x}{\sigma} \right)^{2}} } &
 \vpad{ 1 - e^{-\frac{1}{2} \left( \frac{x}{\sigma} \right)^{2}} } &
 \vpad{ \sigma \sqrt{-2 \ln(1-p)} } &
 \vpad{ \sigma \frac{1}{1-p} \biggl[ \frac{-1}{2 \ln(1-p)} \biggr]^{\frac{1}{2}} } &
 \vpad{ \sigma \frac{1}{n-i+1} \biggl[ \frac{-1}{2 \ln \left( (n-i+1) / n \right)} \biggr]^{\frac{1}{2}} } \\
\text{uniform} &
 \vpad{ \begin{cases} \frac{1}{b-a} & a \leq x \leq b \\
                      0 & \text{else}
        \end{cases} } &
 \vpad{ \begin{cases} 0 & x \le 0 \\
                      \frac{x-a}{b-a} & a \leq x \leq b \\
                      1 & \text{else}
        \end{cases} } &
 \vpad{ a + (b-a) p } &
 \vpad{ b-a } &
 \vpad{ \frac{b-a}{n+1} } \\
\text{Weibull} &
 \vpad{ \frac{a}{b} \left( \frac{x}{b} \right)^{a-1} e^{-\left( \frac{x}{b} \right)^{a}} } &
 \vpad{ 1 - e^{-\left( \frac{x}{b} \right)^{a}} } &
 \vpad{ b \left[ -\ln(1-p) \right]^{\frac{1}{a}} } &
 \vpad{ \frac{b}{a} \frac{1}{1-p} \biggl[ \frac{-1}{\ln(1-p)} \biggr]^{\frac{a-1}{a}} } &
 \vpad{ \frac{b}{a} \frac{1}{n-i+1} \biggl[ \frac{-1}{\ln \left( (n-i+1) / n \right)} \biggr]^{\frac{a-1}{a}} } \\
\hline
\multicolumn{4}{l}{\vpad[8pt]{
 \text{Note: $ z = ( x - \mu ) / \sigma $ has been used to simplify exponents.}
}}
\end{array}
\end{math}
\end{sidewaystable}

Define the approximation error as the absolute difference between the quantile
estimator and the expected spacing as measured in a large number of draws of
$ n $ points.  We used 100 million trials to reduce the noise of the
measurement, and the default parameters given above.  The plots are on
a log-log scale to better separate the curves and values; the scale does not
reflect a functional relationship.  Plots are made for draws of 25, 75, and
250 points.

\begin{figure} \label{fig:grp1}
\includegraphics[width=0.45\textwidth]{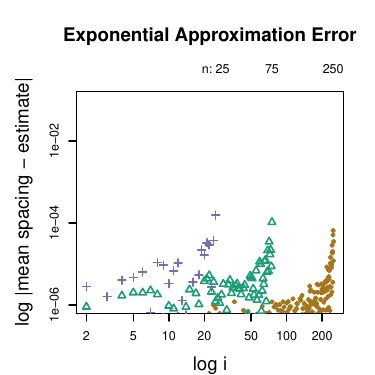} \hfill
\includegraphics[width=0.45\textwidth]{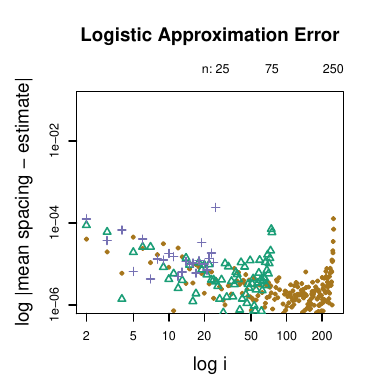} \\
\includegraphics[width=0.45\textwidth]{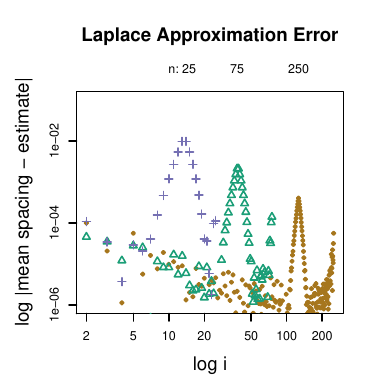} \hfill
\includegraphics[width=0.45\textwidth]{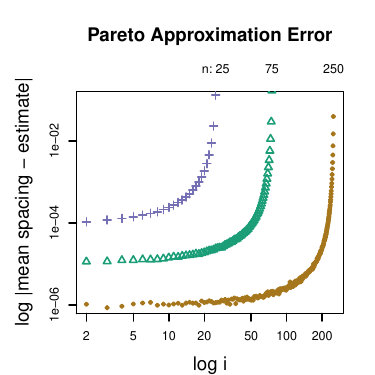} \\
\caption{\label{fig:err.group1}
         Difference between the mean spacing and quantile estimator for the
         $ p $ distributions: the exponential, logistic, Laplace, and Pareto.}
\end{figure}

Figure~\ref{fig:err.group1} presents the first ($ p $) group of distributions.
Because the exponential and logistic estimators are exact, the top two
graphs show the noise floor of the simulations, on the order of
$ 10^{-6} - 10^{-5} $.  The error increases in the tails (one-sided for the
exponential at large $ i $, both large and small $ i $ for the logistic) to
$ 10^{-4} $.  The logarithmic compression on the $ x $ axis seems to create
an imbalance between the left and right tails for the logistic, but the
increase in the error is the same, and matches on the right the behavior of
the exponential.

The error for the Laplace spacing increases for points near the center of
the distribution ($ i = n/2 $).  It is here at the mid-point that the
distribution pastes together two exponentials.  The derivative does not
exist and the actual expected spacing rounds off the joint, producing the
larger error.  The tails follow the error in the logistic results and match
the noise floor.

The one-sided nature of the Pareto distribution shows in its approximation
error, which increases in the tail to the right, for $ i > n/3 $.  The
estimator trails the mean by up to 18\%.  The error at small $ i $ rises
above the noise floor for the smaller draws.  It seems like a correction to
the estimator might be possible, but none could be found.

\begin{figure} \label{fig:grp2}
\includegraphics[width=0.45\textwidth]{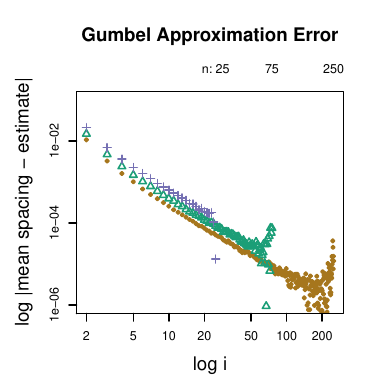} \hfill
\includegraphics[width=0.45\textwidth]{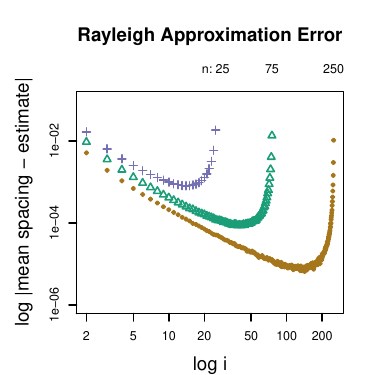} \\
\includegraphics[width=0.45\textwidth]{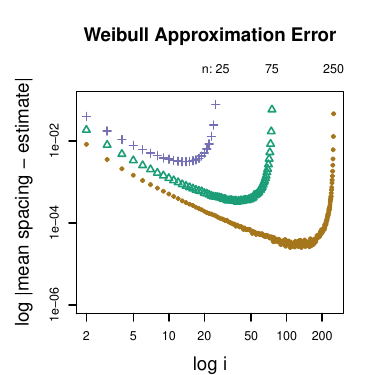} \hfill
\includegraphics[width=0.45\textwidth]{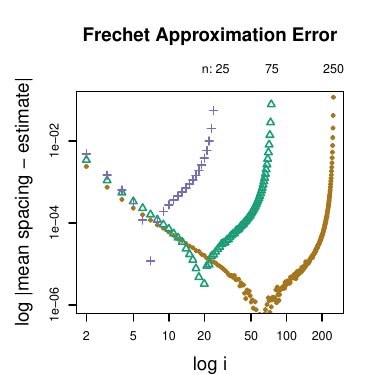} \\
\caption{\label{fig:err.group2}
         Difference between the mean spacing and quantile estimator for the
         $ p $, $ \ln p $ distributions: the Gumbel, Rayleigh, Weibull,
         and Frechet.}
\end{figure}

Figure~\ref{fig:err.group2} shows the approximation error in the second
group of distributions.  Similar to the Pareto, the minimum error shifts
with the draw size, is higher than the noise floor, and changes with the
variate.  The asymmetry of the distribution pushes the minimum a little off
the mid-point at $ n/2 $.  The error rises in the tail, to 6--7\% for the
Gumbel, 13\% for the Rayleigh at $ i = 2 $ and 4--5\% at $ i = n $, and
11\% for the Weibull at small $ i $ and 2--4\% at large.  In all three cases
the estimator is greater than the measured spacing.

The Frechet curves are fundamentally different.  The sharp minimum near $ n/4 $
represents a crossing of the estimator to the measured value.  For small $ i $
the estimator is larger than the measurement, and for large $ i $ it is
smaller.  The error is 8\% at $ i = 2 $ and 26\% at $ i = n $.

\begin{figure}
\begin{minipage}[c]{0.45\linewidth}
 \centering
 \includegraphics{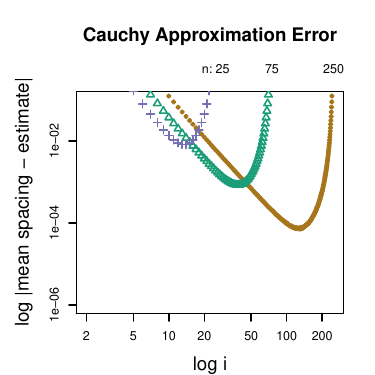} \hfill \\
 \captionof{figure}{
   \label{fig:err.group3}
   Difference between the mean spacing and quantile estimator for the
   other distribution: the Cauchy.}
\end{minipage} \hfill
\begin{minipage}[c]{0.45\linewidth}
 \centering
 \captionof{table}{\label{tbl:minerr} Minimum of the Approximation Error}
 \medskip
 \begin{math}
 \begin{array}{lcccc}
 \hline
 \multicolumn{1}{l}{\vpad{ \text{} }} & \hgap &
 \multicolumn{1}{c}{\vpad{ \text{location } (i) }} & \hgap &
 \multicolumn{1}{c}{\vpad{ \text{value} }} \\
 \hline
 \text{Cauchy} & & $ 0.53 n $ & & \vpad{ 7.2894 / n^{2} } \\
 \text{Pareto} & & 2 & & \vpad{ 0.0949 / n^{2} } \\
 \text{Rayleigh} & & $ 0.56 n $ & & \vpad{ 0.6158 / n^{2} } \\
 \text{Weibull} & & $ 0.51 n $ & & \vpad{ 2.3607 / n^{2} } \\
 \hline
 \end{array}
 \end{math}
\end{minipage}
\end{figure}

The Cauchy approximation error, presented in Figure~\ref{fig:err.group3}, has
a similar form as the Group 2 curves, but increases more rapidly because the
tails are larger.  The minimum error occurs at $ i = n/2 $, since the
distribution is symmetric, and is much larger than the others.

It is clear from these graphs that the minimum error scales inversely with
the square of the size of the draw ($ 1 / n^{2} $).  The proportionality
constant varies over two orders of magnitue (Table~\ref{tbl:minerr}).
The position is near the center of the draw, except for the one-side Pareto.
These results are based on draws of 10 to 250 samples.  The minimum errors
are generally one or two orders of magnitude above the noise floor.

\section{Conclusion}
\label{sec:conclude}

To the known spacing for uniform and exponential variates, we have added
analytical results for draws from a logistic or Gumbel distribution.  The
expected spacing and its variance for the logistic have series whose terms
almost exactly cancel and counteract the combinatorial scaling factor.
This requires using high-precision math libraries, but the results match
numeric integration of the base equations and simulated draws.

The quantile estimator of the expected spacing happens to match the series
for the logistic distribution, and is significantly simpler.  For other
distributions with analytic expressions for the inverse c.d.f. the estimator
is only an approximation.  It is most accurate for middle points, with errors
of $ 10^{-3} $ -- $ 10^{-2} $ for the smallest sample sizes ($ n \leq 25 $),
improving to $ 10^{-5} $ -- $ 10^{-4} $ for large draws ($ n \geq 100 $).
This is about 10 times higher than the noise floor of our simulations.  The
error grows in the tail, and can reach 15--20\% in draws farthest from the
peak of an asymmetric distribution.


\bibliographystyle{amsplain}
\bibliography{spacing}

\renewcommand{\thesection}{Appendix \arabic{section}}
\setcounter{section}{0}


\pagebreak

\section{Spacing for Uniform Variates}
\label{app:unif}

\renewcommand{\theequation}{U.\arabic{equation}}
\setcounter{equation}{0}
\setcounter{table}{0}

\subsubsection*{Density Function}
The density function is
\begin{align*}
f_{D_{i,unif}}(y)
 & = \frac{n!}{(i-2)! (n-i)!} \int_{-\infty}^{\infty} \left( \frac{x-a}{b-a} \right)^{i-2} \left( 1 - \frac{x+y-a}{b-a} \right)^{n-i} \left( \frac{1}{b-a} \right) \left( \frac{1}{b-a} \right) dx \\
 & = \frac{n!}{(i-2)! (n-i)!} \left( \frac{1}{b-a} \right)^{i} \int_{-\infty}^{\infty} \left( \frac{b-x-y}{b-a} \right)^{n-i} \left( x-a \right)^{i-2} dx \\
 & = \frac{n!}{(i-2)! (n-i)!} \left( \frac{1}{b-a} \right)^{n} \int_{a}^{b-y} (x-a)^{i-2} (b-y-x)^{n-i} dx
\end{align*}
where the lower bound of the integral is determined by the larger of $ a $ and
$ a - y $, since the two density functions are 0 below these limits, and the
upper by the smaller of $ b $ and $ b - y $.  From \cite[(3.196.3)]{grad}
(please note that variables in any cited equation may be changed to avoid
conflicts with our analysis)
\[
\int_{\alpha}^{\beta}(x-\alpha)^{\mu-1} (\beta-x)^{\nu-1} dx =
 (\beta-\alpha)^{\mu+\nu-1} B(\mu, \nu)
\]
With $ \mu = i-1 $ and $ \nu = n-i+1 $, both of which are positive and meet
the requirements for the definite integral, and $ \alpha = a $, $ \beta = b-y $,
\[
f_{D_{i,unif}}(y)
 = \frac{n!}{(i-2)! (n-i)!} \left( \frac{1}{b-a} \right)^{n} (b-y-a)^{n-1} B(i-1, n-i+1)
\]
Because the index and sample sizes are integers,
\[
B(x,y) = \frac{\Gamma(x) \Gamma(y)}{\Gamma(x+y)} = \frac{(x-1)! (y-1)!}{(x+y-1)!}
\]
So
\begin{align} \label{appeq:fdi.unif}
f_{D_{i,unif}}(y)
 & = \frac{n!}{(i-2)! (n-i)!} \left( \frac{1}{b-a} \right)^{n} (b-y-a)^{n-1} \frac{(i-2)! (n-i)!}{(n-1)!} \nonumber \\
 & = n \left( \frac{1}{b-a} \right)^{n} (b-y-a)^{n-1} \nonumber \\
 & = \frac{n}{b-a} \left( \frac{b-y-a}{b-a} \right)^{n-1}
\end{align}
For the standard range, $ a=0 $ and $ b=1 $ and this simplifies to
\cite[(2.3)]{pyke65}

\subsubsection*{Expected Spacing}
The expected spacing or first moment is
\begin{align*}
E\bigl\{ D_{i,unif} \bigr\}
 & = \int_{0}^{\infty} y f_{D_{i}}(y) dy \\
 & = \int_{0}^{b-a} y \frac{n}{b-a} \left( \frac{b-a-y}{b-a} \right)^{n-1} dy \\
 & = \frac{n}{(b-a)^{n}} \int_{0}^{b-a} y (b-a-y)^{n-1} dy
\end{align*}
The integral's upper bound is the maximum spacing possible.  Also using
\cite[(3.196.3)]{grad} with $ \mu = 2 $ and $ \nu = n $,
\begin{align} \label{appeq:edi.unif}
E\bigl\{ D_{i,unif} \bigr\}
 & = \frac{n}{(b-a)^{n}} (b-a)^{n+1} B(2, n) \nonumber \\
 & = n (b-a) \frac{1! (n-1)!}{(n+1)!} \nonumber \\
 & = \frac{b-a}{n+1}
\end{align}
The result again matches the formula in Pyke for the unit range.

\subsubsection*{Variance of Spacing}
The variance of the spacing starts with the second moment.  Once again with
\cite[(3.196.3)]{grad} and $ \mu = 3 $ and $ \nu = n $,
\begin{align} \label{appeq:edisq.unif}
E\bigl\{ D_{i,unif}^{2} \bigr\}
 & = \int_{-\infty}^{\infty} y^{2} f_{D_{i}}(y) dy \nonumber \\
 & = \frac{n}{(b-a)^{n}} \int_{0}^{b-a} y^{2} (b-a-y)^{n-1} dy \nonumber \\
 & = \frac{n}{(b-a)^{n}} (b-a)^{n+2} \frac{2! (n-1)!}{(n+2)!} \nonumber \\
 & = \frac{2 (b-a)^{2}}{(n+2) (n+1)}
\end{align}
The variance is
\begin{align} \label{appeq:var.unif}
V\bigl\{ D_{i,unif} \bigr\} & = E\bigl\{ D_{i,unif}^{2} \bigr\}
  - E^{2}\bigl\{ D_{i,unif} \bigr\} \nonumber \\
 & = \frac{2 (b-a)^{2}}{(n+2) (n+1)} - \frac{(b-a)^{2}}{(n+1)^{2}} \nonumber \\
 & = \frac{2 (n+1) - (n+2)}{(n+2) (n+1)^{2}} (b-a)^{2} \nonumber \\
 & = \frac{n}{(n+2)} \left( \frac{b-a}{n+1} \right)^{2}
\end{align}
This too matches Pyke.


\pagebreak

\section{Spacing for Exponential Variates}
\label{app:exp}

\renewcommand{\theequation}{E.\arabic{equation}}
\setcounter{equation}{0}

\subsubsection*{Density Function}
The density function is
\begin{align*}
f_{D_{i,exp}}(y)
 & = \frac{n!}{(i-2)! (n-i)!} \int_{0}^{\infty} \left( 1 - e^{-\lambda x} \right)^{i-2} \left\{ 1 - \left( 1 - e^{-\lambda (x+y)} \right) \right\}^{n-i} \lambda e^{-\lambda x} \lambda e^{-\lambda (x+y)} dx \\
 & = \frac{n!}{(i-2)! (n-i)!} \lambda^{2} \int_{0}^{\infty} \left( 1 - e^{-\lambda x} \right)^{i-2} e^{-\lambda (n-i) x} e^{-\lambda (n-i) y} e^{-\lambda 2x} e^{-\lambda y} dx \\
 & = \frac{n!}{(i-2)! (n-i)!} \lambda^{2} e^{-\lambda (n-i+1) y} \int_{0}^{\infty} \left( 1 - e^{-\lambda x} \right)^{i-2} e^{-\lambda (n-i+2) x} dx
\end{align*}
where the restriction $ x \ge 0 $ sets the lower limit.  From
\cite[(3.312.1)]{grad}
\begin{equation*}
 \int_0^{\infty} \left( 1 - e^{-x/\beta} \right)^{\nu - 1} e^{-\mu x} dx
  = \beta B(\beta \mu, \nu) 
  = \beta \frac{(\beta \mu - 1)! (\nu - 1)!}{(\beta \mu + \nu - 1)!}
\end{equation*}
With $ \beta = 1/\lambda $, $ \nu = i-1 $, $ \mu = \lambda (n-i+2) $,
$ \beta \mu = n-i+2 $
\begin{align} \label{appeq:fdi.exp}
f_{D_{i,exp}}(y)
 & = \frac{n!}{(i-2)! (n-i)!} \lambda^{2} e^{-\lambda (n-i+1)}
  \frac{1}{\lambda} \frac{(n-i+1)! (i-2)!}{n!} \nonumber \\
 & = \lambda (n-i+1) e^{-\lambda (n-i+1) y}
\end{align}
This is one factor of \cite[(2.9)]{pyke65}.

\subsubsection*{Expected Spacing}
The expected spacing is
\begin{equation*}
E\bigl\{ D_{i,exp} \bigr\}
 = \int_{0}^{\infty} y f_{D_{i}}(y) dy 
 = \lambda (n-i+1) \int_{0}^{\infty} y e^{-\lambda (n-i+1) y} dy
\end{equation*}
Using \cite[(2.322.1)]{grad}
\begin{equation*}
 \int x e^{ax} dx = e^{ax} \left( \frac{x}{a} - \frac{1}{a^2} \right)
\end{equation*}
we get
\begin{align} \label{appeq:edi.exp}
E\bigl\{ D_{i,exp} \bigr\}
 & = \lambda (n-i+1) \left[ e^{-\lambda (n-i+1) y}
  \left\{ \frac{y}{-\lambda (n-i+1)} - \frac{1}{\lambda^{2} (n-i+1)^{2}} \right\} \right]_{0}^{\infty} \nonumber \\
 & = \left[ -y e^{-\lambda (n-i+1) y} -
       \frac{1}{\lambda (n-i+1)} e^{-\lambda (n-i+1) y}
     \right]_{0}^{\infty} \nonumber \\
 & = \frac{1}{\lambda (n-i+1)}
\end{align}
where the exponential dominates at the upper bound, driving both terms to
zero, and the first term drops at the lower bound.

Some authors define a normalized spacing for exponential variates, multiplying
\eqref{appeq:edi.exp} by $ (n - i + 1) $ to give a constant expected value of
$ 1 / \lambda $; see \cite{proschanpyke} for an example.  This compenstates
for the spacing's growth in the tails.  Note that the factor also appears in
many of the quantile estimators, often in combination with a second scaling
factor for the other tail.  The second factor sometimes involves $ i - 1 $
and sometimes $ \ln n - i + 1 $ or $ \ln i - 1 $.

\subsubsection*{Variance of Spacing}
Following what we did for the uniform distribution, we calculate first the
second moment,
\[
E\bigl\{ D_{i,exp}^2 \bigr\}
 = \int_{0}^{\infty} y^{2} f_{D_{i}}(y) dy
 = \lambda (n-i+1) \int_{0}^{\infty} y^{2} e^{-\lambda (n-i+1) y} dy
\]
with \cite[(2.322.2)]{grad}
\[
\int x^{2} e^{ax} dx
 = e^{ax} \left( \frac{x^{2}}{a} - \frac{2x}{a^{2}} + \frac{2}{a^{3}} \right)
\]
$ a = -\lambda (n-i+1) $ is also the factor before the integral, leaving
\begin{align} \label{appeq:edisq.exp}
E\bigl\{ D_{i,exp}^2 \bigr\}
 & = \left[ e^{-\lambda (n-i+1) y} \left\{ -y^{2} - \frac{2y}{\lambda (n-i+1)}
           - \frac{2}{\lambda^{2} (n-i+1)^{2}} \right\} \right]_{0}^{\infty}
           \nonumber \\
 & = \frac{2}{\lambda^{2} (n-i+1)^{2}}
\end{align}
Here too only the last term survives at the lower bound.  The variance then
follows as
\begin{align} \label{appeq:var.exp}
V\bigl\{ D_{i,exp} \bigr\}
 & = E\bigl\{ D_{i,exp}^{2} \bigr\} - E^{2}\bigl\{ D_{i,exp} \bigr\} \nonumber \\
 & = \frac{2}{\lambda^{2} (n-i+1)^{2}} - \frac{1}{\lambda^{2} (n-i+1)^{2}} \nonumber \\
 & = \frac{1}{\lambda^{2} (n-i+1)^{2}}
\end{align}


\pagebreak

\section{Spacing for Logistic Variates}
\label{app:logis}

\renewcommand{\theequation}{L.\arabic{equation}}
\setcounter{equation}{0}

\subsubsection*{Density Function}
Starting with the distribution's density functions and substituting
$ z = (x - \mu) / \sigma $, $ dz = dx / \sigma $, and $ w = y / \sigma $,
\begin{align*}
f(x) & = \frac{e^{-z}}{\sigma (1+e^{-z})^{2}} 
 & f(x+y) & = \frac{e^{-z} e^{-w}}{\sigma (1 + e^{-z} e^{-w})^{2}} \\
F(x) & = \frac{1}{1 + e^{-z}} 
 & F(x+y) & = \frac{1}{1 + e^{-z} e^{-w}}
\end{align*}
The spacing's density function is
\begin{align} \label{appeq:fdi.log.mid}
f_{D_{i,logis}}(y)
 & = \frac{n!}{(i-2)! (n-i)!} \int_{-\infty}^{\infty} \left( \frac{1}{1+e^{-z}} \right)^{i-2} \left(1 - \frac{1}{1 + e^{-z} e^{-w}} \right)^{n-i} \left( \frac{e^{-z}}{\sigma (1+e^{-z})^{2}} \right) \left( \frac{e^{-z} e^{-w}}{\sigma (1 + e^{-z} e^{-w})^{2}} \right) \sigma dz \nonumber \\
 & = \frac{n!}{(i-2)! (n-i)!} \frac{1}{\sigma} e^{-w} \int_{-\infty}^{\infty} \left( \frac{1}{1+e^{-z}} \right)^{i-2} \left( \frac{e^{-z} e^{-w}}{1 + e^{-z} e^{-w}} \right)^{n-i} e^{-2z} \left( \frac{1}{1+e^{-z}} \right)^{2} \left( \frac{1}{1+e^{-z}e^{-w}} \right)^{2} dz \nonumber \\
 & = \frac{n!}{(i-2)! (n-i)!} \frac{1}{\sigma} e^{-w} \int_{-\infty}^{\infty} \left( \frac{1}{1+e^{-z}} \right)^{i} \left( \frac{1}{1+e^{-z}e^{-w}} \right)^{n-i+2} e^{-(n-i)z} e^{-(n-i)w} e^{-2z} dz \nonumber \\
 & = \frac{n!}{(i-2)! (n-i)!} \frac{1}{\sigma} e^{-(n-i+1)w} \int_{-\infty}^{\infty} \left( \frac{1}{1+e^{-z}} \right)^{i} \left( \frac{1}{1+e^{-z}e^{-w}} \right)^{n-i+2} e^{-(n-i+2)z} dz \nonumber \\
 & = \frac{n!}{(i-2)! (n-i)!} \frac{1}{\sigma} e^{-(n-i+1)w} \int_{-\infty}^{\infty} \left( \frac{1}{1+e^{-z}} \right)^{i} \left( \frac{e^{w}}{e^{w}+e^{-z}} \right)^{n-i+2} e^{-(n-i+2)z} dz \nonumber \\
 & = \frac{n!}{(i-2)! (n-i)!} \frac{1}{\sigma} e^{w} \int_{-\infty}^{\infty} \left( \frac{1}{1+e^{-z}} \right)^{i} \left( \frac{1}{e^{w}+e^{-z}} \right)^{n-i+2} e^{-(n-i+2)z} dz
\end{align}
This has the form of the definite integral \cite[(3.315.1)]{grad}
\begin{equation*}
\int_{-\infty}^{\infty} \left[ \frac{1}{e^{\beta} + e^{-x}} \right]^{\nu} \left[ \frac{1}{e^{\gamma} + e^{-x}} \right]^{\rho} e^{-\mu x} dx = 
 e^{(\mu - \rho) \gamma - \beta \nu} B(\mu, \nu+\rho-\mu) {}_{2}F_{1}\left( \nu, \mu; \nu+\rho; 1-e^{\gamma-\beta} \right)
\end{equation*}
We have $ \beta = 0 $, $ \gamma = w $, $ \rho = n-i+2 $, $ \mu = n-i+2 $, and
$ \nu = i $.  These values satisfy the conditions on the integral:
$ \lvert \imag(\beta) \rvert = 0 < \pi $;
$ \lvert \imag(\gamma) \rvert = 0 < \pi $; and
$ \real(\nu + \rho) = n+2 > \real(\mu) = n-i+2 > 0 $.  Now
$ (\mu - \rho) \gamma - \beta \nu $ is zero and the exponential falls out,
leaving
\begin{align} \label{appeq:fdi.log}
f_{D_{i,logis}}(y)
 & = \frac{n!}{(i-2)! (n-i)!} \frac{1}{\sigma} e^{w} B(n-i+2,i)
   {}_{2}F_{1}\left( i, n-i+2; n+2; 1-e^{w} \right) \nonumber \\
 & = \frac{1}{\sigma} e^{w} \frac{n!}{(i-2)! (n-i)!} \frac{(n-i+1)! (i-1)!}{(n+1)!}
   {}_{2}F_{1}\left( i, n-i+2; n+2; 1-e^{w} \right) \nonumber \\
 & = \frac{1}{\sigma} e^{y/\sigma} \frac{(n-i+1) (i-1)}{n+1}
   {}_{2}F_{1}\left( i, n-i+2; n+2; 1-e^{y/\sigma} \right)
\end{align}

\subsubsection*{Expected Spacing}
To avoid integrating the hypergeometric function we can change the order of
the integrals when calculating the expected spacing.  Starting with
\eqref{appeq:fdi.log.mid} and making the same substitutions $ z $ and $ w $,
so $ dw = dy / \sigma $ and the integration bounds stay the same,
\begin{align*}
E\bigl\{ D_{i,logis} \bigr\}
 & = \int_{0}^{\infty} y f_{D_{i}}(y) dy \\
 & = \frac{n!}{(i-2)! (n-i)!}
 \int_{0}^{\infty} \frac{y}{\sigma} e^{w} dy \int_{-\infty}^{\infty} \left( \frac{1}{1+e^{-z}} \right)^{i} \left( \frac{1}{e^{w}+e^{-z}} \right)^{n-i+2} e^{-(n-i+2)z} dz \\
 & = \frac{\sigma n!}{(i-2)! (n-i)!} \int_{-\infty}^{\infty} \left( \frac{1}{1+e^{-z}} \right)^{i} e^{-(n-i+2)z} \int_{0}^{\infty} w e^{w} \left( \frac{1}{e^{w}+e^{-z}} \right)^{n-i+2} dw dz
\end{align*}
Doing the inner integral by parts,
\[
I_{in} = \int_{0}^{\infty} w e^{w} \left( \frac{1}{e^{w} + e^{-z}} \right)^{n-i+2} dw
\]
we set $ u = w $, $ du = dw $, and
\[
dv = e^{w} \left( \frac{1}{e^{w} + e^{-z}} \right)^{n-i+2} dw
\]
Letting $ \eta = n-i+2 $, $ t = e^{w} + e^{-z} $, and
$ dt = e^{w} dw = (t - e^{-z}) dw $,
\begin{align*}
dv & = t^{-\eta} dt \\
v & = \frac{t^{-(\eta-1)}}{-(\eta-1)}
\end{align*}
and the integral becomes
\begin{align*}
I_{in}
 & = \bigl[ uv \bigr]_{0}^{\infty} - \int_{0}^{\infty} v du \\
 & = \left[ \frac{w}{-(\eta-1)} \left( \frac{1}{e^{w}+e^{-z}} \right)^{\eta-1} \right]_{0}^{\infty}
  - \frac{1}{-(\eta-1)} \int_{0}^{\infty} \left( \frac{1}{e^{w}+e^{-z}} \right)^{\eta-1} d w \\
\intertext{The first term goes to zero at both limits, so}
I_{in}
 & = \frac{1}{\eta-1} \int_{0}^{\infty} \left( \frac{1}{t} \right)^{\eta-1}
   \frac{1}{t-e^{-z}} dt
\end{align*}
The integral has a known solution \cite[(2.117.4)]{grad}
\[
\int \frac{dx}{x^{m} (a+bx)} =
 \frac{(-1)^{m} b^{m-1}}{a^{m}} \ln \left( \frac{a+bx}{x} \right)
  + \sum_{k=1}^{m-1} \frac{(-1)^{k} b^{k-1}}{(m-k) a^{k} x^{m-k}}
\]
With $ m = \eta - 1 $, $ a = -e^{-z} $, and $ b = 1 $ this becomes
\begin{align*}
\int \left( \frac{1}{t} \right)^{\eta-1} \frac{dt}{t-e^{-z}} 
 & = \frac{(-1)^{\eta-1} (1)^{\eta-2}}{\left( -e^{-z} \right)^{\eta-1}} \ln \left( \frac{-e^{-z}+t}{t} \right) 
  + \sum_{k=1}^{\eta-2} \frac{(-1)^{k} (1)^{k-1}}{(\eta-1-k)\left(-e^{-z}\right)^{k}} \left( \frac{1}{t} \right)^{\eta-1-k} \\
  & = \frac{1}{e^{-(\eta-1)z}} \ln \left( \frac{e^{w}}{e^{w}+e^{-z}} \right)
   + \sum_{k=1}^{\eta-2} \frac{1}{\eta-1-k} \frac{1}{e^{-kz}} \left( \frac{1}{e^{w}+e^{-z}} \right)^{\eta-1-k} \\
  & = e^{(\eta-1) z} \ln\left( \frac{e^{w}}{e^{w}+e^{-z}} \right)
   + \sum_{k=1}^{\eta-2} \frac{1}{\eta-1-k} e^{kz} \left( \frac{1}{e^{w}+e^{-z}} \right)^{\eta-1-k} \\
\end{align*}
The series is ignored if $ n = i $.  At $ w = \infty $ the first term goes to
$ \ln(1) $, which is zero, and the
exponential in the second term's denominator drives it to zero, because the
exponent $ \eta-1-k $ is positive.  At $ w = 0 $ the exponentials go to one,
leaving
\[
I_{in} = \frac{-1}{\eta-1} \left\{ -e^{(\eta-1)z} \ln\left( 1+e^{-z} \right)
 + \sum_{k=1}^{\eta-2} \frac{1}{\eta-1-k} e^{kz} \left( \frac{1}{1+e^{-z}} \right)^{\eta-1-k} \right\}
\]
If we substitute back $ \eta $ we have finally
\begin{equation}
I_{in}
 = \frac{1}{n-i+1} \left\{
  e^{(n-i+1)z} \ln\left( 1+e^{-z} \right)
  - \sum_{k=1}^{n-i} \frac{1}{n-i+1-k} e^{kz} \left( \frac{1}{1+e^{-z}} \right)^{n-i+1-k} \right\}
\end{equation}
Evaluating the outer integral,
\begin{align*}
E\bigl\{ D{_i,logis} \bigr\}
 & = \frac{\sigma n!}{(i-2)! (n-i)!} \int_{-\infty}^{\infty} \left( \frac{1}{1+e^{-z}} \right)^{i} e^{-(n-i+2)z} I_{in} dz \\
 & = \frac{\sigma n!}{(i-2)! (n-i+1)!} \int_{-\infty}^{\infty} e^{-z} \left( \frac{1}{1+e^{-z}} \right)^{i} \ln\left( 1+e^{-z} \right) dz \\
 & \qquad - \frac{\sigma n!}{(i-2)! (n-i+1)!} \sum_{k=1}^{n-i} \frac{1}{n-i+1-k} \int_{-\infty}^{\infty} e^{-(n-i+2-k)z} \left( \frac{1}{1+e^{-z}} \right)^{n+1-k} dz
\end{align*}
To evaluate the first integral, transform $ t = 1 + e^{-z} $ and
$ dt = -e^{-z} dz $, which changes the integration limits, giving
\[
\int_{-\infty}^{\infty} e^{-z} \left( \frac{1}{1+e^{-z}} \right)^{i} \ln\left( 1 + e^{-z} \right) dz = \int_{\infty}^{1} -\left( \frac{1}{t} \right)^{i} \ln t \: dt = \int_{1}^{\infty} t^{-i} \ln t \: dt
\]
which we integrate by parts with $ u = \ln t $, $ du = dt / t $,
$ dv = t^{-i} dt $, and $ v = t^{-i+1} / (-i + 1) $.  This means
\begin{align*}
\int_{1}^{\infty} t^{-i} \ln t \: dt
 & = \left[ \frac{1}{-i+1} t^{-i+1} \ln t \right]_{1}^{\infty} - \int_{1}^{\infty} \frac{1}{-i+1} t^{-i+1} \frac{1}{t} dt \\
 & = \left[ \frac{1}{-i+1} t^{-i+1} \ln t \right]_{1}^{\infty} - \int_{1}^{\infty} \frac{1}{-i+1} t^{-i} dt \\
 & = \left[ \frac{1}{-i+1} t^{-i+1} \ln t \right]_{1}^{\infty} - \left[ \left( \frac{1}{-i+1} \right)^{2} t^{-i+1} \right]_{1}^{\infty} \\
 & = \left[ \frac{1}{-i+1} t^{-i+1} \ln t - \frac{1}{(-i+1)^2} t^{-i+1}\right]_{1}^{\infty} \\
 & = \left( \frac{1}{-i+1} \right)^{2}
\end{align*}
because at the upper limit the $ t^{-i+1} $ goes to 0 (remember, $ i >= 2 $),
and at the lower the $ t $ dependence disappears.

The second integral requires another known formula, \cite[(3.314)]{grad}
\[
\int_{-\infty}^{\infty} \frac{e^{-\mu x}}{\left( e^{\beta/\gamma} + e^{-x/\gamma} \right)^{\nu}} dx = \gamma e^{\beta(\mu - \frac{\nu}{\gamma})} B(\gamma \mu, \nu - \gamma \mu)
\]
We have $ \mu = n-i+2-k $, $ \beta = 0 $, $ \gamma = 1 $, and $ \nu = n+1-k $.
The integral has a number of conditions, all of which are met:
$ \lvert \imag(\beta) \rvert = 0 < \pi \real(\gamma) = \pi $;
$ \real(\nu/\gamma) = n+1-k > \real(\mu)  = n-i+2-k$, which holds since
$ i \ge 2 $; and $ \real(\mu) = n-i+2-k > 0 $, which is true at the largest
$ k = n-i $.  With $ \beta = 0 $ the exponential vanishes and the integral is
\[
\int_{-\infty}^{\infty} e^{-(n-i+2-k)z} \left( \frac{1}{1+e^{-z}} \right)^{n+1-k} dz = B(n-i+2-k, i-1)
\]

Combining the two and expanding the beta function, then dividing out the
$ n-i+1-k $ factor before the second integral, gives
\begin{equation} \label{appeq:edi.log}
E\bigl\{ D_{i,logis} \bigr\}
 = \frac{\sigma n!}{(i-2)! (n-i+1)!}
   \left\{ \left( \frac{1}{i-1} \right)^{2}
    - \sum_{k=1}^{n-i} \frac{(n-i-k)! (i-2)!}{(n-k)!} \right\}
\end{equation}

\subsubsection*{Variance of Spacing}
We will use the same approach to calculate $ E\bigl\{ D^{2}_{i,logis} \bigr\} $.
\begin{align} \label{appeq:edisq1}
E\bigl\{ D^{2}_{i,logis} \bigr\}
 & = \frac{n!}{(i-2)! (n-i)!}
    \int_{-\infty}^{\infty} \left\{ F(x) \right\}^{i-2} f(x)
    \int_{0}^{\infty} y^{2} \left\{1 - F(x+y)\right\}^{n-i} f(x+y) dy dx
    \nonumber \\
 & = \frac{n!}{(i-2)! (n-i)!}
      \int_{-\infty}^{\infty} \left( \frac{1}{1+e^{-z}} \right)^{i} e^{-(n-i+2) z} dz
      \int_{0}^{\infty} \frac{y^{2}}{\sigma} e^{y/\sigma}
        \left( \frac{1}{e^{y/\sigma} + e^{-z}} \right)^{n-i+2} dy
\end{align}
Let $ w = y / \sigma $ so $ dw = dy / \sigma $ and the integration limits
don't change, and again set $ \eta = n - i + 2 $.  Then
\begin{align}
E\bigl\{ D^{2}_{i,logis} \bigr\}
 & = \frac{\sigma^{2} n!}{(i-2)! (n-i)!}
      \int_{-\infty}^{\infty} \left( \frac{1}{1+e^{-z}} \right)^{i} e^{-\eta z} dz
      \int_{0}^{\infty} w^{2} e^{w} \left( \frac{1}{e^{w} + e^{-z}} \right)^{\eta} dw \nonumber \\
 & = \frac{\sigma^{2} n!}{(i-2)! (n-i)!}
      \int_{-\infty}^{\infty} \left( \frac{1}{1+e^{-z}} \right)^{i} e^{-\eta z} I_{in}(z) dz \\
I_{in}(z) & = \int_{0}^{\infty} w^{2} e^{w} \left( \frac{1}{e^{w} + e^{-z}} \right)^{\eta} dw \nonumber
\end{align}

The inner integral solves by parts with $ u = w^{2} $ and $ dv $ the
remainder.  Let $ t = e^{w} + e^{-z} $, $ dt = e^{w} dw $ (and note
$ dt = (t - e^{-z}) dw $ for later).  Then the second part is
\begin{equation*}
v = \int t^{-\eta} dt = \frac{1}{-(\eta - 1)} t^{-(\eta - 1)}
 = \frac{1}{-(\eta - 1)} \left( \frac{1}{e^{w} + e^{-z}} \right)^{\eta - 1}
\end{equation*}
Assembling,
\begin{align*}
I_{in}(z)
 & = uv \biggr\rvert_{0}^{\infty} - \int_{0}^{\infty} v du \\
 & = \left[ \frac{w^{2}}{-(\eta-1)} \left( \frac{1}{e^{w}+e{-z}} \right)^{\eta-1} \right]_{0}^{\infty}
    - \int_{0}^{\infty} \frac{2w}{-(\eta-1)} \left( \frac{1}{e^{w}+e^{-z}} \right)^{\eta-1} dw
\end{align*}
The first term at $ w = 0 $ is 0, and at $ \infty $ the exponential factor and
positive $ \eta - 1 $ dominate and drive it to 0; the first term drops out.  We
solve the second term again by parts.  With the same $ t $ substitution
\begin{equation*}
dv = \left( \frac{1}{e^{w} + e^{-z}} \right)^{\eta-1} dw 
   = \left( \frac{1}{t} \right)^{\eta-1} \frac{dt}{t - e^{-z}}
\end{equation*}
This has the form of a known indefinite integral (\cite[(2.117.4)]{grad})
\begin{equation} \label{appeq:2.117.4}
\int \frac{dx}{x^{m} (a+bx)}
 = \frac{(-1)^{m} b^{m-1}}{a^{m}} \ln\left(\frac{a+bx}{x}\right)
  + \sum_{k=1}^{m-1} \frac{-1^{k} b^{k-1}}{(m-k) a^{k} x^{m-k}}
\end{equation}
with $ m = \eta - 1 $, $ a = -e^{-z} $, and $ b = 1 $.  So
\begin{equation*}
v =
 \sum_{k=1}^{\eta-2} \frac{1}{\eta-1-k} \frac{1}{e^{-k z}}
   \left( \frac{1}{e^{w}+e^{-z}} \right)^{\eta-1-k}
 + \frac{1}{e^{-(\eta-1) z}} \ln\frac{e^{w}}{e^{w}+e^{-z}}
\end{equation*}
where we ignore the sum if $ \eta = 2 $, i.e. $ n = i $.  Note that in this case
the integral is equivalent to \cite[(2.118.1)]{grad}.  Assembling
\begin{align*}
I_{in}(z)
 & = uv \biggr\rvert_{0}^{\infty} - \int_{0}^{\infty} v du \\
 & = \frac{2}{\eta-1} \left\{ 
   \begin{aligned}
   \left[
     \sum_{k=1}^{\eta-2} \frac{1}{\eta-1-k} e^{k z} w
       \left( \frac{1}{e^{w}+e^{-z}} \right)^{\eta-1-k}
     w e^{(\eta-1) z} \ln\frac{e^{w}}{e^{w}+e^{-z}}
   \right]_{0}^{\infty} \\
    - \int_{0}^{\infty} \sum_{k=1}^{\eta-2} \frac{1}{\eta-1-k} e^{kz} \left( \frac{1}{e^{w}+e^{-z}} \right)^{\eta-1-k} dw
    - \int_{0}^{\infty} e^{(\eta-1) z} \ln \frac{e^{w}}{e^{w}+e^{-z}} dw 
  \end{aligned} \right\}
\end{align*}
At $ w = \infty $ the $ uv $ term goes to 0; this also happens at $ w = 0 $.
This leaves the two integrals.  In the second we split out the $ k = \eta - 2 $
term because it solves with a different definite integral (and as we said, if
$ \eta = 2 $ the sum is ignored).
\begin{align*}
I_{in}(z) &
 = \frac{-2}{\eta-1}
  \left\{ \begin{aligned}
    \int_{0}^{\infty} e^{(\eta-2) z} \left( \frac{1}{e^{w}+e^{-z}} \right) dw \\
    + \int_{0}^{\infty} \sum_{k=1}^{\eta-3} \frac{1}{\eta-1-k} e^{kz} \left( \frac{1}{e^{w}+e^{-z}} \right)^{\eta-1-k} dw \\
    + \int_{0}^{\infty} e^{(\eta-1) z} \ln\frac{e^{w}}{e^{w}+e^{-z}} dw
  \end{aligned} \right\} \\
 & = \frac{-2}{\eta-1} \left\{ I_{in1} + I_{in2} + I_{in3} \right\}
\end{align*}

The first integral uses the same $ t $ substitution and \cite[(2.118.1)]{grad}
\begin{equation} \label{appeq:2.118.1}
\int \frac{dx}{x (a+bx)} = - \frac{1}{a} \ln(\frac{a+bx}{x})
\end{equation}
with $ a = -e^{-z} $ and $ b = 1 $.
\begin{align} \label{appeq:in2}
I_{in1}
 & = \int_{0}^{\infty} e^{(\eta-2) z} \left( \frac{1}{e^{w}+e^{-z}} \right)^{\eta-1-k} dw \nonumber \\
 & = \int_{0}^{e^{-z}} \frac{dt}{t^{\eta-1-k}} (t-e^{-z}) \nonumber \\
 & = e^{(\eta-2) z} frac{-1}{-e^{-z}} \ln\frac{t-e^{-z}}{t} \biggr\rvert_{0}^{-e^{-z}}
   \nonumber \\
 & = e^{(\eta-2) z} e^{z} \ln\frac{e^{w}}{e^{w}+e^{-z}} \biggr\rvert_{0}^{\infty} 
   \nonumber \\
 & = e^{(\eta-1) z} \ln(1+e^{-z})
\end{align}
In the last step, at $ w = \infty $ the logarithm goes to 0, and at $ w = 0 $
it becomes $ -ln(1+e^{-z}) $.

The second integral again involves \cite[(2.117.4)]{grad} after substituting
$ t = e^{w} + e^{-z} $, $ dt = e^{w} dw = (t - e^{-z}) dw $.
\begin{align} \label{appeq:in3}
I_{in2}
 & = \int_{0}^{\infty} \sum_{k=1}^{\eta-3} \frac{1}{\eta-1-k} e^{kz}
      \left( \frac{1}{e^{w}+e^{-z}} \right)^{\eta-1-k} dw \nonumber \\
 & = \sum_{k=1}^{\eta-3} \frac{1}{\eta-1-k} e^{kz}
      \int_{0}^{\infty} \left( \frac{1}{t} \right)^{\eta-1-k} \frac{dt}{t-e^{-z}} \nonumber \\
\intertext{With $ m = \eta - 1 - k $, $ a = -e^{-z} $, and $ b = 1 $
 in \eqref{appeq:2.117.4}}
 & = \sum_{k=1}^{\eta-3} \frac{1}{\eta-1-k} e^{kz} \left\{
      \frac{(-1)^{\eta-1-k} (1)^{\eta-2-k}}{(-e^{-z})^{\eta-1-k}} \ln\frac{t-e^{-z}}{t}
      + \sum_{l=1}^{\eta-2-k} \frac{(-1)^{l} (1)^{l-1}}{(\eta-1-k-l) e^{-lz}}
         \left( \frac{1}{e^{w}+e^{-z}} \right)^{\eta-1-k-l} \right\}_{0}^{\infty}
         \nonumber \\
 & = \sum_{k=1}^{\eta-3} \frac{1}{\eta-1-k} e^{kz} \left\{
      e^{(\eta-1-k) z} \ln\frac{e^{w}}{e^{w}+e^{-z}}
      + \sum_{l=1}^{\eta-2-k} \frac{1}{\eta-1-k-l} e^{lz}
         \left( \frac{1}{e^{w}+e^{-z}} \right)^{\eta-1-k-l} \right\}_{0}^{\infty}
         \nonumber \\
\intertext{where both terms go to zero at $ w = \infty $, leaving}
 & = \sum_{k=1}^{\eta-3} \frac{1}{\eta-1-k} e^{kz} \left\{
     e^{(\eta-1-k) z} \ln(1+e^{-z}) - \sum_{l=1}^{\eta-2-k} \frac{1}{\eta-1-k-l} e^{lz} \left( \frac{1}{1+e^{-z}} \right)^{\eta-1-k-l} \right\}
\end{align}

For the third of these we substitute $t = e^{-z-w}$,
$ dt = -e^{-z-w} dw = -t dw $, which transforms the integration limits to
$ e^{-z} $ and 0.
\begin{align*}
I_{in3}
 & = \int_{0}^{\infty} e^{(\eta-1) z} \ln\frac{e^{w}}{e^{w}+e^{-z}} dw \\
 & = -e^{(\eta-1) z} \int_{0}^{\infty} \ln\left(1+e^{-zw}\right) dw \\
 & = -e^{(\eta-1) z} \int_{0}^{e^{-z}} \frac{1}{t} \ln(1+t) dt
\end{align*}
This has no known integral and we must switch to the power series of the
logarithm, integrating term by term.  Using \cite[(1.511)]{grad}
\begin{equation*}
\ln(1+x) = \sum_{k=1}^{\infty} (-1)^{k+1} \frac{x^{k}}{k}
\end{equation*}
the integral becomes
\begin{align} \label{appeq:in1}
I_{in3}
 & = -e^{(\eta-1) z} \int_{0}^{e^{-z}} \frac{1}{t}
       \sum_{k=1}^{\infty} (-1)^{k+1} \frac{t^{k}}{k} dt \nonumber \\
 & = -e^{(\eta-1) z} \sum_{k=1}^{\infty} (-1){k-1}
       \int_{0}^{e^{-z}} \frac{t^{k-1}}{k} dt \nonumber \\
 & = -e^{(\eta-1) z} \sum_{k=1}^{\infty} (-1){k-1}
       \left[ \frac{t^{k}}{k^2} \right]_0^{e^{-z}} \nonumber \\
 & = -e^{(\eta-1) z} \sum_{k=1}^{\infty} (-1){k-1} \frac{e^{-zk}}{k^2} \nonumber \\
 & = e^{(\eta-1) z} \sum_{k=1}^{\infty} \frac{\left(-e^{-z}\right)^{k}}{k^2} \nonumber \\
 & = e^{(\eta-1) z} \li2(-e^{-z})
\end{align}
where $\li2()$ is the dilogarithm.  It converges within the unit circle.  When
we integrate over $ z $ we will need a transformation outside the circle.

Combining all three integrals we get
\begin{align} \label{appeq:in}
I_{in}(z)
 & = \frac{-2}{\eta-1} \left\{ \begin{aligned}
   e^{(\eta-1) z} \ln(1+e^{-z}) \\
   + \sum_{k=1}^{\eta-3} \frac{1}{\eta-1-k} e^{kz}
       \left\{ e^{(\eta-1-k) z} \ln(1+e^{-z})
        - \sum_{l=1}^{\eta-2-k} \frac{1}{\eta-1-k-l} e^{lz}
          \left( \frac{1}{1+e^{-z}} \right)^{\eta-1-k-l} \right\} \\
   + e^{(\eta-1) z} \li2(-e^{-z})
    \end{aligned} \right\} \nonumber \\
 & = \frac{2}{\eta-1} \left\{ \begin{aligned}
   - e^{(\eta-1) z} \ln(1+e^{-z})
   - \sum_{k=1}^{\eta-3} \frac{1}{\eta-1-k} e^{(\eta-1) z} \ln(1+e^{-z}) \\
   + \sum_{k=1}^{\eta-3} \sum_{l=1}^{\eta-2-k}
       \frac{1}{\eta-1-k} \frac{1}{\eta-1-k-l} e^{(k+l)z}
         \left( \frac{1}{1+e^{-z}} \right)^{\eta-1-k-l} \\
   - e^{(\eta-1) z} \li2(-e^{-z})
   \end{aligned} \right\} \nonumber \\
 & = \frac{2}{\eta-1} \left\{ \begin{aligned}
   - \sum_{k=1}^{\eta-2} \frac{1}{\eta-1-k} e^{(\eta-1) z} \ln(1+e^{-z}) \\
   + \sum_{k=1}^{\eta-3} \sum_{l=1}^{\eta-2-k}
       \frac{1}{\eta-1-k} \frac{1}{\eta-1-k-l} e^{(k+l) z}
         \left( \frac{1}{1+e^{-z}} \right)^{\eta-1-k-l} \\
   - e^{(\eta-1) z} \li2(-e^{-z})
   \end{aligned} \right\}
\end{align}
The first term in the second step is the same as the case $ k = \eta-2 $ in
the first sum, so we have combined the two in the third step by shifting the
upper limit.  As usual, ignore the series if the upper bound is less than the
lower: for $ i = n $ ($ \eta = 2 $) in the first series, or $ i = n-1 $ or
$ i =n $ in the second.

Now we can evaluate the outer integral, \eqref{appeq:edisq1}.
\begin{align*} 
E\bigl\{ D_{i,logis}^{2} \bigr\}
 & = \frac{\sigma^{2} n!}{(i-2)! (n-i)!} \int_{-\infty}^{\infty}
     \left( \frac{1}{1+e^{-z}} \right)^{i} e^{-\eta z}
     \frac{-2}{\eta-1} I_{in} dz \nonumber \\
 & = \frac{2 \sigma^{2} n!}{(i-2)! (n-i+1)!} \int_{-\infty}^{\infty}
     \left( \frac{1}{1+e^{-z}} \right)^{i} e^{-\eta z} I_{in} \nonumber \\
 & = \frac{2 \sigma^{2} n!}{(i-2)! (n-i+1)!}
  \left\{ \begin{aligned}
    - \sum_{k=1}^{\eta-2} \frac{1}{\eta-1-k} \int_{-\infty}^{\infty}
      e^{-z} \left( \frac{1}{1+e^{-z}} \right)^{i} \ln(1+e^{-z}) dz \\
    + \sum_{k=1}^{\eta-3} \sum_{l=1}^{\eta-2-k} \frac{1}{\eta-1-k} \frac{1}{\eta-1-k-l}
     \int_{-\infty}^{\infty} e^{(k+l-\eta) z} \left( \frac{1}{1+e^{-z}} \right)^{n+1-k-l} dz \\
    - \int_{-\infty}^{\infty} e^{-z} \left( \frac{1}{1+e^{-z}} \right)^{i} \li2(-e^{-z}) dz
  \end{aligned} \right\} \nonumber \\
\end{align*}
\begin{align} \label{appeq:edisq2}
E\bigl\{ D_{i,logis}^{2} \bigr\}
 & = \frac{2 \sigma^{2} n!}{(i-2)! (n-i+1)!}
  \left\{ \begin{aligned}
    - \sum_{k=1}^{\eta-2} \frac{1}{\eta-1-k} I_{out1} \\
    + \sum_{k=1}^{\eta-3} \sum_{l=1}^{\eta-2-k} \frac{1}{\eta-1-k} \frac{1}{\eta-1-k-l} I_{out2} \\
    - I_{out3}
  \end{aligned} \right\} \\
\end{align}

After substituting $ t = 1 + e^{-z} $, $ dt = - e^{-z} dz $ and changing the
integration bounds $ z = -\infty $ to $ t = \infty $ and $ z = \infty $ to
$ t = 1 $, the first outer integral becomes
\begin{align*}
I_{out1}
 & = \int_{-\infty}^{\infty}
       e^{-z} \left( \frac{1}{1+e^{-z}} \right)^{i} \ln(1+e^{-z}) dz \nonumber \\
 & = \int_{1}^{\infty} t^{-i} \ln t dt
\end{align*}
This integral also appears in the outer integral for the expected spacing.
Integrating by parts with $ u = \ln t $, $ du = dt / t $, $ dv = t^{-i} dt $,
and $ v = t^{-i+1} / (-i+1) $ we have
\begin{align} \label{appeq:out1}
I_{out1}
 & = \left[ \frac{1}{-i+1} t^{-i+1} \ln t \right]_{1}^{\infty}
   - \int_{1}^{\infty} \frac{1}{-i+1} t^{-i+1} \frac{1}{t} dt \nonumber \\
 & = \left[ \frac{1}{-i+1} t^{-i+1} \ln t \right]_{1}^{\infty}
   - \int_{1}^{\infty} \frac{1}{-i} t^{-i} dt \nonumber \\
 & = \left[ \frac{1}{-i+1} t^{-i+1} \ln t \right]_{1}^{\infty}
   - \left[ \frac{1}{-i+1} \frac{1}{-i+1} t^{-i+1} \right]_{1}^{\infty} \nonumber \\
 & = \left[ \frac{t^{-i+1}}{-i+1} \ln t - \frac{t^{-i+1}}{(-i+2)^2} \right]_{1}^{\infty} \nonumber \\
 & = \left( \frac{1}{-i+1} \right)^{2}
\end{align}

The second integral is known \cite[(3.314)]{grad}
\begin{equation} \label{appeq:3.314}
\int_{-\infty}^{\infty} \frac{e^{-\mu x}}{(e^{\beta/\gamma} + e^{-x/\gamma})^\nu} dx =
 \gamma e^{\beta (\mu - \nu/\gamma)} B(\gamma \mu, \nu - \gamma \mu)
\end{equation}
where $ B() $ is the beta function, which for integral arguments simplifies to
\begin{equation} \label{appeq:beta}
B(a,b) = \frac{(a-1)! (b-1)!}{(a+b-1)!}
\end{equation}
Fitting our integral, we have $ \mu = \eta-k-l $, $ \nu = n+1-k-l $,
$ \beta = 0 $, and $ \gamma = 1 $.  The definite integral has a number of
conditions to check: $ \lvert \beta \rvert < \pi \real(\gamma) $, i.e.
$ 0 < \pi $ ; $ \real(\nu/\gamma) > \real(\mu) $, or
$ n-i+2-k-l > n+1-k-l $ which is true because $ i \ge 2 $; and
$ \real(\mu) > 0 $, or $ n-i+2-k-l > 0 $ with either of the maximum indices
$ k_{max} = n-i-1 $, $ l = 1 $ or $ k = 1 $, $ l_{max} = n-i $.  Then
\begin{align} \label{appeq:out2}
I_{out2}
 & = \int_{-\infty}^{\infty} e^{-(\eta-k-l) z} \left( \frac{1}{1+e^{-z}} \right)^{n+1-k-l} dz \nonumber \\
 & = B(\eta-k-l, n+1-k-l) \nonumber \\
 & = B(n-i+2-k-l, n+1-k-l) \nonumber \\
 & = \frac{(n-i+1-k-l)! (i-2)!}{(n-i+2-k-l+(i-1)-1)!} \nonumber \\
 & = \frac{(\eta-1-k-l)! (i-2)!}{(n-k-l)!}
\end{align}

The dilogarithm integral must be split to stay within its radius of
convergence, the unit circle.  This occurs at $ z = 0 $; that is, for
$ z \ge 0 $ we are within the unit circle, and $ z < 0 $ outside.
We use the transformation \cite[(3.2)]{dilog}
\begin{equation} \label{appeq:invdilog}
\li2(\frac{1}{z}) = - \li2(z) - \frac{\pi^2}{6} - \frac{1}{2} (\ln(-z))^2
\end{equation}
In other words, we end up calculating
\begin{align} \label{appeq:out3}
I_{out3}
 & = \int_{-\infty}^{\infty} e^{-z} \left( \frac{1}{1+e^{-z}} \right)^{i} \li2(-e^{-z}) dz \nonumber \\
 & = \int_{0}^{\infty} e^{-z} \left( \frac{1}{1+e^{-z}} \right)^{i} \li2(-e^{-z}) dz
   - \int_{-\infty}^{0} e^{-z} \left( \frac{1}{1+e^{-z}} \right)^{i} \li2(-e^{z}) dz \nonumber \\
 & \qquad - \int_{-\infty}^{0} e^{-z} \left( \frac{1}{1+e^{-z}} \right)^{i} \frac{\pi^{2}}{6} dz
   - \int_{-\infty}^{0} e^{-z} \left( \frac{1}{1+e^{-z}} \right)^{i} \frac{1}{2} (\ln(e^{z}))^{2} dz \nonumber \\
 & = \int_{0}^{\infty} e^{-z} \left( \frac{1}{1+e^{-z}} \right)^{i} \sum_{k=1}^{\infty} \frac{(-e^{-z})^{k}}{k^2} dz 
  - \int_{-\infty}^{0} e^{-z} \left( \frac{1}{1+e^{-z}} \right)^{i} \sum_{k=1}^{\infty} \frac{(-e^{z})^{k}}{k^2} dz \nonumber \\
 & \qquad  - \frac{\pi^2}{6} \int_{-\infty}^{0} e^{-z} \left( \frac{1}{1+e^{-z}} \right)^{i} dz 
   - \frac{1}{2} \int_{-\infty}^{0} z^{2} e^{-z} \left( \frac{1}{1+e^{-z}} \right)^{i} dz
 \nonumber \\
  & = \sum_{k=1}^{\infty} \frac{(-1)^k}{k^2} \int_{0}^{\infty} e^{-(k+1) z} \left( \frac{1}{1+e^{-z}} \right)^{i} dz
   - \sum_{k=1}^{\infty} \frac{(-1)^k}{k^2} \int_{0}^{\infty} e^{(k-1) z} \left( \frac{1}{1+e^{-z}} \right)^{i} dz \nonumber \\
 & \qquad   - \frac{\pi^2}{6} \int_{-\infty}^{0} e^{-z} \left( \frac{1}{1+e^{-z}} \right)^{i} dz
   - \frac{1}{2} \int_{-\infty}^{0} z^{2} e^{-z} \left( \frac{1}{1+e^{-z}} \right)^{i} dz \nonumber \\
  & = \sum_{k=1}^{\infty} \frac{(-1)^{k}}{k^2} I_{out3A}
   - \sum_{k=1}^{\infty} \frac{(-1)^{k}}{k^2} I_{out3B}
   - \frac{\pi^2}{6} I_{out3C}
   - \frac{1}{2} I_{out3D}
\end{align}

The first two series integrals we tackle the same way.  By substituting
$ t = e^{-z} $, $ dt = -e^{-z} dz $ we convert each to a rational function
that must be solved repeatedly by parts until the degree is reduced to one,
at which point a simple integral finishes the process.

The first integral converts to
\begin{align*}
I_{out3A} & = \int_{0}^{\infty} e^{-(k+1) z} \left( \frac{1}{1+e^{-z}} \right)^{i} dz \\
 & = \int_{0}^{1} t^{k} \left( \frac{1}{1+t} \right)^{i} dt
\end{align*}
This fits the form of the \cite[(2.111)]{grad} solutions.  The starting point
is \cite[(2.111.2)]{grad}
\begin{equation} \label{appeq:2.111.2}
\int \frac{x^j dx}{(a+bx)^{m}} =
 \frac{x^{j}}{(a+bx)^{m-1} (j+1-m) b} -
 \frac{j a}{(j+1-m) b} \int \frac{x^{j-1} dx}{(a+bx)^{m}}
\end{equation}
with $ j = k $ counting down, $ m = i $, and $ a = b = 1 $.  This obviously
fails if $ j = m-1 $, so if we start below this value the recursion ends at
\cite[(2.111.1)]{grad}
\begin{equation} \label{appeq:2.111.1}
\int \frac{dx}{(a+bx)^{m}} = \frac{1}{b (-m+1)} \frac{1}{(a+bx)^{m-1}}
\end{equation}
If we do not, then another recursion takes over, \cite[(2.111.3)]{grad}
\begin{equation} \label{appeq:2.111.3}
\int \frac{x^{m-1} dx}{(a+bx)^{m}} =
 - \frac{x^{m-1}}{(a+bx)^{m-1} (m-1) b}
 + \frac{1}{b} \int \frac{x^{m-2} dx}{(a+bx)^{m-1}}
\end{equation}
Both exponents decrease by one in the next step, so the form continues to
hold and we continue down this chain.  It ends at \cite[(2.111.1, second
formula)]{grad}
\begin{equation} \label{appeq:2.111.1b}
\int \frac{dx}{a+bx} = \frac{1}{b} \ln(a+bx)
\end{equation}

If we start with $ k < i-1 $, the recursion becomes
\begin{equation*}
I_{out3A} = \sum_{j=1}^{k} \frac{t^{j}}{(1+t)^{i-1}} \frac{-1}{i-j-1} \prod_{l=j+1}^{k} \frac{l}{i - 1 - l} - \frac{1}{i-1} \frac{1}{(1+t)^{i-1}} \prod_{l=1}^{k} \frac{l}{i-1-l}
\end{equation*}
The first sum comes from \eqref{appeq:2.111.2}, the last term from
\eqref{appeq:2.111.1}.  Here we've inverted the signs so that the denominator
factors are always positive.  The $ j+1 $ lower bound on the product in
the sum represents the scaling from the previous step; as usual, when the
lower bound is greater than the upper for the $ j = k $ step, the product
is ignored, or considered to be one.  For a given $ j $ the product can
re-written with factorials to represent the cut-off of factors.
\begin{equation*}
\frac{k (k-1) (k-2) \ldots (j+1)}{(i-k-1) (i-k) (i-k+1) \ldots (i-(j+1)-1)}
 = \frac{k!}{(l_{min}-1)!} \frac{(i-(l_{max}-1)-1)!}{(i-(j-1)-1)!}
 = \frac{k!}{j!} \frac{(i-k-2)!}{(i-j-2)!}
\end{equation*}
For the \eqref{appeq:2.111.1} term this also applies, with $ l_{min} = 1 $ or
$ j = 0 $, so that product becomes
\begin{equation*}
\frac{k! (i-k-2)!}{(i-2)!}
\end{equation*}
Thus,
\begin{equation*}
I_{out3A} = \sum_{j=1}^{k} \frac{t^{j}}{(1+t)^{i-1}} \frac{-1}{i-j-1} \frac{k!}{j!} \frac{(i-k-2)!}{(i-j-2)!} - \frac{1}{(1+t)^{i-1}} \frac{1}{i-1} \frac{k! (i-k-2)!}{(i-2)!}
\end{equation*}
At the integration bounds $ t=1 $ and $ t = 0 $, this simplifies to
\begin{align} \label{appeq:out3a1}
I_{out3A} & =
 \sum_{j=1}^{k} \frac{1}{2^{i-1}} \frac{-1}{i-j-1} \frac{k!}{j!} \frac{(i-k-2)!}{(i-j-2)!}
 - \frac{1}{2^{i-1}} \frac{1}{i-1} \frac{k! (i-k-2)!}{(i-2)!}
 + \frac{1}{i-1} \frac{k! (i-k-2)!}{(i-2)!} \nonumber \\
 & = \sum_{j=1}^{k} \frac{1}{2^{i-1}} \frac{-1}{i-j-1} \frac{k!}{j!} \frac{(i-k-2)!}{(i-j-2)!}
 + \frac{1}{i-1} \left( 1 - \frac{1}{2^{i-1}} \right) \frac{k! (i-k-2)!}{(i-2)!}
 \nonumber \\
 & = \frac{k! (i-k-2)!}{2^{i-1}} \left\{ \frac{2^{i-1}-1}{(i-1)!} - \sum_{j=1}^{k} \frac{1}{j! (i-j-1)!} \right\}
\end{align}

If $ k \ge i - 1 $ then we have a different set of series for the integral.
\begin{multline*}
I_{out3A} = \sum_{j=i}^{k} \frac{t^{j}}{(1+t)^{i-1}} \frac{(-1)^{k-j}}{j+1-i} \prod_{l=j+1}^{k} \frac{l}{l+1-i} \\
 + \sum_{j=2}^{i} (-1)^{k-i} \left( \frac{t}{1+t} \right)^{j-1} \frac{1}{j-1} \prod_{l=i}^{k} \frac{l}{l+1-i}
 + (-1)^{k-i+1} \ln(1+t) \prod_{l=i}^{k} \frac{l}{l+1-i}
\end{multline*}
The products in the second sum and third term are frozen when we leave the
\eqref{appeq:2.111.2} recursion for \eqref{appeq:2.111.3}.  The first series
represents that first chain, as can be seen with the constant $ i-1 $ exponent
in the denominator, and the second series the other, with the same exponent
for the $ t $ and $ t+1 $ factors.  The product is now
\begin{equation*}
\frac{k (k-1) (k-2) \ldots (j+1)}{(k+1-i) (k-i) (k-1-i) \ldots (j+2-i)}
 = \frac{k!}{(l_{min}-1)!} \frac{(l_{min} - i)!}{(k+1-i)!}
 = \frac{k!}{j!} \frac{(j+1-i)!}{(k+1-i)!}
\end{equation*}
with $ l_{min} = j + 1 $.  The frozen value is
\begin{equation*}
\frac{k!}{(i-1)! (k+1-i)!} = \binom{k}{i-1}
\end{equation*}
Shifting the index of the second sum by one, the indefinite integral is
\begin{align*}
I_{out3A} & = \sum_{j=i}^{k} \frac{t^{j}}{(1+t)^{i-1}} \frac{(-1)^{k-j}}{j+1-i}
 \frac{k!}{j!} \frac{(j+1-i)!}{(k+1-i)!} \\
 & \qquad + (-1)^{k-i} \sum_{j=1}^{i-1} \left( \frac{t}{1+t} \right)^{j} \frac{1}{j} \binom{k}{i-1}
 + (-1)^{k+1-i} \binom{k}{i-1} \ln(1+t)
\end{align*}
Evaluating at the integration bound $ t = 0 $ everything drops, leaving for
$ t = 1 $
\begin{align} \label{appeq:out3a2}
I_{out3A} & 
 = \sum_{j=i}^{k} \frac{1}{2^{i-1}} \frac{(-1)^{k-j}}{j+1-i} \frac{k!}{j!}
     \frac{(j+1-i)!}{(k+1-i)!}
 + (-1)^{k-i} \binom{k}{i-1} \sum_{j=1}^{i-1} \frac{1}{j} \frac{1}{2^{j}}
 + (-1)^{k+1-i} \binom{k}{i-1} \ln 2 \nonumber \\
 & = \sum_{j=i}^{k} \frac{1}{2^{i-1}} \frac{(-1)^{k-j}}{j+1-i} \frac{k!}{j!}
     \frac{(j+1-i)!}{(k+1-i)!}
   + (-1)^{k-i} \binom{k}{i-1} \left\{ \sum_{j=1}^{i-1} \frac{1}{j} \frac{1}{2^{j}} - \ln 2 \right\} \nonumber \\
 & = (-1)^{k-1} \frac{k!}{(k+1-i)!} \left[
  \sum_{j=i}^{k} \frac{(-1)^{j}}{2^{i-1}} \frac{(j-i)!}{j!}
  + (-1)^{i} \frac{1}{(i-1)!} \left\{ \sum_{j=1}^{i-1} \frac{1}{j 2^{j}} - \ln 2
  \right\} \right]
\end{align}
The second sum has a clear interpretation.  With \cite[(0.241)]{grad}
\begin{equation*}
\sum_{k=1}^{\infty} \frac{1}{k} \frac{1}{2^{k}} = \ln 2
\end{equation*}
we see that it is equivalent to the infinite sum starting at $ i $
\begin{equation*}
\sum_{j=1}^{i-1} \frac{1}{j 2^{j}} - \ln 2 =
 - \sum_{j=i}^{\infty} \frac{1}{j 2^{j}}
\end{equation*}

The second integral, with the same substitution, has the opposite sign
to the $ t $ exponent.
\begin{align*}
I_{out3B} & = \int_{-\infty}^{0} e^{(k-1) z} \left( \frac{1}{1+e^{-z}} \right)^{i} dz \\
 & = \int_{1}^{\infty} \left( \frac{1}{t} \right)^{k} \left( \frac{1}{1+t} \right)^{i} dt
\end{align*}
This follows the \cite[(2.117)]{grad} indefinite integrals.  Begin by
decreasing the $ t $ exponent until it reaches one \cite[(2.117.1)]{grad}.
\begin{equation} \label{appeq:2.117.1}
\int \frac{dx}{x^{n} (a+bx)^{m}} = \frac{-1}{(n-1) a x^{n-1} (a+bx)^{m-1}}
 + \frac{b (2-n-m)}{a (n-1)} \int \frac{dx}{x^{n-1} (a+bx)^{m}}
\end{equation}
Then change the $ 1+t $ exponent until it reaches one \cite[(2.117.3)]{grad}.
\begin{equation} \label{appeq:2.117.3}
\int \frac{dx}{x (a+bx)^m} = \frac{1}{a (m-1) (a+bx)^{m-1}}
 + \frac{1}{a} \int \frac{dx}{x (a+bx)^{m-1}}
 \end{equation}
Finally terminate the chain with \eqref{appeq:2.118.1}.  For these integrals
$ a = b = 1 $, the \eqref{appeq:2.117.1} sum is over the dilogarithm index
$ k $, and the \eqref{appeq:2.117.3} sum over the spacing index $ i $.
Writing the recursions as series,
\begin{align*}
I_{out3B} & = \sum_{j=2}^{k} \frac{-1}{(j-1) t^{j-1} (1+t)^{i-1}}
 \prod_{l=j+1}^{k} \frac{2-l-i}{l-1} 
 + \sum_{j=2}{i} \frac{1}{(j-1) (1+t)^{j-1}} \prod_{l=1}^{k} \frac{2-l-i}{l-1} \\
 & \qquad - \ln\frac{1+t}{t} \prod_{l=1}^{k} \frac{2-l-i}{l-1} \\
 & =  \sum_{j=2}^{k} \frac{(-1)^{k-j+1}}{(j-1) t^{j} (1+t)^{i-1}} \prod_{l=j+1}^{k} \frac{i+l-2}{l-1}
 + \sum_{j=2}^{i} \frac{(-1)^{k-1}}{(j-1) (1+t)^{j-1}} \prod_{l=1}^{k} \frac{l+i-2}{l-1} \\
 & \qquad - (-1)^{k-1} \ln \frac{1+t}{t} \prod_{l=1}^{k} \frac{l+i-2}{l-1}
\end{align*}
The product expands to
\begin{equation*}
\frac{(i+k-2) (i+k-3) (i+k-4) \ldots (i+j-1)}{(k-1) (k-2) (k-3) \ldots j}
 = \frac{(i+k-2)!}{(i+j-2)!} \frac{(j-1)!}{(k-1)!}
\end{equation*}
When finished with \eqref{appeq:2.117.1}, the scaling factor fixes at $ l = 2 $
or $ j = 1 $,
\begin{equation*}
\frac{(i+k-2)!}{(i-1)! (k-1)!}
\end{equation*}
So
\begin{align*}
I_{out3B} & = \sum_{j=2}^{k} \frac{(-1)^{k-j+1}}{(j-1) t^{j} (1+t)^{i-1}}
 \frac{(i+k-2)!}{(i+j-2)!} \frac{(j-1)!}{(k-1)!} \\
 & \qquad + \frac{(i+k-2)!}{(i-1)! (k-1)!} (-1)^{k-1} \left\{
   \sum_{j=2}^{i} \frac{1}{(j-1) (1+t)^{j-1}} - \ln \frac{1+t}{t} \right\}
\end{align*}
At the integration bound $ t = \infty $ the factors of $ t $ in both series
are in the denominator, so they go to zero.  The log term is also 0.  This
leaves only the $ t = 1 $ bound, or
\begin{align} \label{appeq:out3b}
I_{out3B} & =
 - \sum_{j=2}^{k} \frac{(-1)^{k-j+1}}{(j-1) 2^{i-1}}
   \frac{(i+k-2)!}{(i+j-2)!} \frac{(j-1)!}{(k-1)!}
 - \frac{(i+k-2)!}{(i-1)! (k-1)!} (-1)^{k-1} \left\{
   \sum_{j=2}^{i} \frac{1}{(j-1) 2^{j-1}} - \ln 2 \right\} \nonumber \\
 & = \frac{(-1)^{k} (i+k-2)!}{(k-1)!} \left[
  \sum_{j=2}^{k} \frac{(-1)^{j}}{2^{i-1}} \frac{(j-2)!}{(i+j-2)!}
  + \frac{1}{(i-1)!} \left\{ \sum_{j=1}^{i-1} \frac{1}{j 2^{j}} - \ln 2
  \right\} \right]
\end{align}

The third integral goes directly.  Let $ t = 1 + e^{-z} $, $ dt = -e^{-z} dz $.
\begin{align} \label{appeq:out3c}
I_{out3C} & = \int_{-\infty}^{0} e^{-z} \left( \frac{1}{1+e^{-z}} \right)^{i} dz \nonumber \\
 & = \int_{2}^{\infty} t^{-i} dt \nonumber \\
 & = \frac{1}{-i+1} t^{-i+1} \biggr\rvert_{2}^{\infty} \nonumber \\
 & = - \frac{2^{-i+1}}{-i+1}
\end{align}

The fourth integral is done by parts, with $ u = z^{2} $, $ du = 2z dz $,
and $ dv = e^{-z} (1/(1+e^{-z}))^{i} dz $.  This integrates directly by
substituting $ t = 1 + e^{-z} $, $ dt = -e^{-z} dz $,
\begin{equation*}
v = - \int t^{-i} dt = \frac{-1}{-i+1} \left( \frac{1}{1+e^{-z}} \right)^{i-1}
\end{equation*}
At both integration bounds the product $ u v $ disappears, from the $ u $
contribution at $ t = 0 $, from the reciprocal exponential in $ v $ at
infinity.
\begin{align*}
I_{out3D} & = \int_{-\infty}^{0} z^{2} e^{-z} \left( \frac{1}{1+e^{-z}} \right)^{i} dz \\
 & = uv \biggr\rvert_{-\infty}^{0} - \int_{-\infty}^{0} v du \\
 & = \left[ z^{2} \frac{-1}{-i+1} \left( \frac{1}{1+e^{-z}} \right)^{i-1} \right]_{-\infty}^{0}
  - \int_{-\infty}^{0} \frac{-1}{-i+1} \left( \frac{1}{1+e^{-z}} \right)^{i-1} 2z dz \\
  & = \int_{-\infty}^{0} \frac{2}{-i+1} z \left( \frac{1}{1+e^{-z}} \right)^{i-1} dz
\end{align*}
Again working by parts, $ u = z $, $ du = dz $ , with $ dv $ the remainder.
For $ v $ we use \eqref{appeq:2.117.4} with $ m = i - 1 $, $ a = 1 $, and
$ b = -1 $.
\begin{align*}
v & = \int t^{-i+1} \frac{dt}{1-t} \\
 & = \frac{(-1)^{i-1} (-1)^{i-2}}{(1)^{i-1}} \ln \frac{1-t}{t}
    + \sum_{j=1}^{i-2} \frac{(-1)^{j} (-1)^{j-1}}{(i-1-j) (1)^{j} t^{i-1-j}} \\
 & = - \ln \frac{e^{-z}}{1+e^{-z}}
     - \sum_{j=1}^{i-2} \frac{1}{i-1-j} \left( \frac{1}{1+e^{-z}} \right)^{i-1-j}
\end{align*}
If $ i = 2 $ ignore the series; the integral reduces to \eqref{appeq:2.118.1}.
At the lower integration limit the logarithm drives the first term to zero,
as does the exponential in the denominator the second term.  Both terms
also drop at $ t = 0 $, leaving
\begin{align*}
I_{out3D}
 & = uv \biggr\rvert_{-\infty}^{0} - \int_{-\infty}^{0} v du \\
 & = \frac{2}{-i+1} \left[ -z \ln \frac{e^{-z}}{1+e^{-z}}
           - \sum_{j=1}^{i-2} \frac{z}{i-1-j} \left( \frac{1}{1+e^{-z}} \right)^{i-1-j} \right]_{\-infty}^{0} \\
 & \qquad - \frac{2}{-i+1} \int_{-\infty}^{0} \left\{
      -\ln \frac{e^{-z}}{1+e^{-z}}
      - \sum_{j=1}^{i-2} \frac{1}{i-1-j} \left( \frac{1}{1+e^{-z}} \right)^{i-1-j} \right\} dz \\
 & = \frac{2}{-i+1} \int_{-\infty}^{0} \ln \frac{e^{-z}}{1+e^{-z}} dz
   + \frac{2}{-i+1} \int_{-\infty}^{0} \sum_{j=1}^{i-2} \frac{1}{i-1-j} \left( \frac{1}{1+e^{-z}} \right)^{i-1-j} dz \\
 & = \frac{-2}{-i+1} \int_{-\infty}^{0} \ln (1+e^{z}) dz
   + \frac{2}{-i+1} \sum_{j=1}^{i-2} \frac{1}{i-1-j}
      \int_{-\infty}^{0} \left( \frac{1}{1+e^{-z}} \right)^{i-1-j} dz \\
 & = \frac{-2}{-1+1} \int_{0}^{\infty} ln(1+e^{-z}) dz
   + \frac{2}{-i+1} \sum_{j=1}^{i-2} \frac{1}{i-1-j}
      \int_{-\infty}^{0} \left( \frac{1}{1+e^{-z}} \right)^{i-1-j} dz
\end{align*}
The first integral here has been solved \cite[(4.223.1)]{grad}
\begin{equation} \label{appeq:4.223.1}
\int_{0}^{\infty} \ln(1+e^{-x}) dx = \frac{\pi^2}{12}
\end{equation}
The second integral we've just done in the last integration by parts, with
$ m = i - 1 - j $, $ a = 1 $, and $ b = -1 $, giving
\begin{align*}
\int_{-\infty}^{0} \left( \frac{1}{1+e^{-z}} \right)^{i-1-j} dz
 & = \frac{(-1)^{i+j} (-1)^{i-1-j}}{(1)^{i-1-j}} \ln \frac{e^{-z}}{1+e^{-z}} \\
 & \qquad  + \sum_{l=1}^{i-2-j} \frac{1}{i-1-j-l} \frac{(-1)^{l} (-1)^{l-1}}{(1)^{l}}
     \left( \frac{1}{1+e^{-z}} \right)^{i-1-j-l} \\
 & = - \ln \frac{e^{-z}}{1+e^{-z}}
   - \sum_{l=1}^{i-2-j} \frac{1}{i-1-j-l} \left( \frac{1}{1+e^{-z}} \right)^{i-1-j-l}
\end{align*}
which goes to zero at the $ t = -\infty $ limit, leaving terms at $ z = 0 $.
The final result for the fourth integral is
\begin{equation} \label{appeq:out3d}
I_{out3D} = \frac{2}{-i+1} \left\{ -\frac{\pi^2}{12} +
   \sum_{j=1}^{i-2} \frac{1}{i-1-j} \left[ \ln 2
     - \sum_{l=1}^{i-2-j} \frac{1}{i-1-j-l} \frac{1}{2^{i-1-j-l}} \right] \right\}
\end{equation}

We can pull everything together.  Combining \eqref{appeq:edisq2} and
\eqref{appeq:out3} and then substituting the individual results
\eqref{appeq:out1}, \eqref{appeq:out2}, \eqref{appeq:out3c}, and
\eqref{appeq:out3d}, and removing $ \eta = n - i + 2 $ first in Step 3 by
substituting $ m = \eta - 1 - k $ and then in Step 4 reverting
$ m \rightarrow k $,
\begin{align} \label{appeq:edisq}
E\bigl\{ D_{i,logis}^{2} \bigr\}
 & = \frac{2 \sigma^{2} n!}{(i-2)! (n-i+1)!}
  \left\{ \begin{aligned}
    - \sum_{k=1}^{n-i} \frac{1}{n-i+1-k} I_{out1} \\
    + \sum_{k=1}^{n-i-1} \sum_{l=1}^{n-i-k} \frac{1}{n-i+1-k} \frac{1}{n-i+1-k-l} I_{out2} \\
    - I_{out3}
  \end{aligned} \right\} \nonumber \\
 & = \frac{2 \sigma^{2} n!}{(i-2)! (n-i+1)!}
  \left\{ \begin{aligned}
    - \sum_{k=1}^{n-i} \frac{1}{n-i+1-k} I_{out1} \\
    + \sum_{k=1}^{n-i-1} \sum_{l=1}^{n-i-k} \frac{1}{n-i+1-k} \frac{1}{n-i+1-k-l} I_{out2} \\
    + \sum_{k=1}^{\infty} \frac{(-1)^{k}}{k^2} ( I_{out3B} - I_{out3A} )
    + \frac{\pi^2}{6} I_{out3C}
    + \frac{1}{2} I_{out3D}
  \end{aligned} \right\} \nonumber \\
 & = \frac{2 \sigma^{2} n!}{(i-2)! (n-i+1)!}
  \left\{ \begin{aligned}
    - \sum_{m=1}^{n-i} \frac{1}{m} \left( \frac{1}{-i+1} \right)^{2} \\
    + \sum_{m=2}^{n-i} \sum_{l=1}^{m-1} \frac{1}{m} \frac{1}{m-l} \frac{(m-l)! (i-2)!}{(m-1-l+i)!} \\
    + \sum_{k=1}^{\infty} \frac{(-1)^{k}}{k^2} ( I_{out3B} - I_{out3A} ) \\
    - \frac{\pi^2}{6} \frac{2^{-i+1}}{-i+1} \\
    + \frac{1}{2} \frac{2}{-i+1} \left\{ -\frac{\pi^2}{12} +
       \sum_{j=1}^{i-2} \frac{1}{i-1-j} \left[ \ln 2
     - \sum_{l=1}^{i-2-j} \frac{1}{i-1-j-l} \frac{1}{2^{i-1-j-l}} \right] \right\}
  \end{aligned} \right\} \nonumber \\
 & = \frac{2 \sigma^{2} n!}{(i-2)! (n-i+1)!}
  \left\{ \begin{aligned}
    - \sum_{k=1}^{n-i} \frac{1}{k} \left( \frac{1}{-i+1} \right)^{2} \\
    + \sum_{k=2}^{n-i} \sum_{l=1}^{k-1} \frac{1}{k} \frac{(k-1-l)! (i-2)!}{(k-1-l+i)!} \\
    + \sum_{k=1}^{\infty} \frac{(-1)^{k}}{k^2} ( I_{out3B} - I_{out3A} ) \\
    - \frac{1}{-i+1} \frac{\pi^2}{12} \frac{1}{2^{i}} \\
    + \frac{1}{-i+1} \sum_{j=1}^{i-2} \frac{1}{i-1-j} \left[ \ln 2
     - \sum_{l=1}^{i-2-j} \frac{1}{i-1-j-l} \frac{1}{2^{i-1-j-l}} \right]
  \end{aligned} \right\} \\ \nonumber
\end{align}

The dilogarithm series has not been expanded, because we have to consider
the two cases.  If $ k < i-1 $, using \eqref{appeq:out3b} and
\eqref{appeq:out3a1}
\begin{align} \label{appeq:dilog1}
I_{out3B} - I_{out3A}
 & = \frac{(-1)^{k} (i+k-2)!}{(k-1)!}
   \left[ \sum_{j=2}^{k} \frac{(-1)^{j}}{2^{i-1}} \frac{(j-2)!}{(i+j-2)!}
    + \frac{1}{(i-1)!} \left\{ \sum_{j=1}^{i-1} \frac{1}{j 2^{j}} - \ln 2
    \right\} \right] \nonumber \\
 & \qquad - \frac{k! (i-k-2)!}{2^{i-1}} \left\{
    \frac{2^{i-1} - 1}{(i-1)!} - \sum_{j=1}^{k} \frac{1}{j! (i-j-1)!} \right\}
\end{align}
otherwise with \eqref{appeq:out3b} and \eqref{appeq:out3a2}
\begin{align} \label{appeq:dilog2}
I_{out3B} - I_{out3A}
 & = \frac{(-1)^{k} (i+k-2)!}{(k-1)!}
  \left[ \sum_{j=2}^{k} \frac{(-1)^{j}}{2^{i-1}} \frac{(j-2)!}{(i+j-2)!}
   + \frac{1}{(i-1)!} \left\{ \sum_{j=1}^{i-1} \frac{1}{j 2^{j}} - \ln 2
   \right\} \right] \nonumber \\
 & \qquad - \frac{(-1)^{k} k!}{(k+1-i)!}
  \left[ \sum_{j=1}^{k} \frac{(-1)^{j}}{2^{i-1}} \frac{(j-i)!}{j!}
   + (-1)^{i} \frac{1}{(i-1)!} \left\{
   \sum_{j=1}^{i-1} \frac{1}{j 2^{j}} - \ln 2 \right\} \right]
\end{align}

The variance follows by subtracting the square of the expected spacing.


\pagebreak

\section{Spacing for Gumbel Variates}
\label{app:gumbel}

\renewcommand{\theequation}{G.\arabic{equation}}
\setcounter{equation}{0}

\subsubsection*{Density Function}
Starting with the distribution's density functions and using
$ z = e^{-(x+y-\mu)/\sigma} $, $ dz = -(z / \sigma) dx $, and
$ w = e^{y/\sigma} $, so that $ wz = e^{-(x-\mu)/\sigma} $,
\begin{align*}
f(x) & = \frac{wz}{\sigma} e^{-wz}
 & f(x+y) & = \frac{z}{\sigma} e^{-z} \\
F(x) & = e^{-wz} 
 & F(x+y) & = e^{-z}
\end{align*}
Including $ y $ in the $ z $ substitution simplifies $ F(x+y) $, which in turn
will simplify the form of the spacing's density function.  The integration
limits change from $ x = [-\infty, +\infty] $ to $ z = [+\infty, 0] $.
\begin{align*}
f_{D_{i,gumb}}(y)
 & = \frac{n!}{(i-2)! (n-i)!} \int_{\infty}^{0} \left( e^{-wz} \right)^{i-2}
  \left( 1 - e^{-z} \right)^{n-i} \left( \frac{wz}{\sigma} e^{-wz} \right)
  \left( \frac{z}{\sigma} e^{-z} \right) \left( -\frac{\sigma}{z} dz \right) \\
 & = \frac{n!}{(i-2)! (n-i)!} \frac{w}{\sigma} \int_{0}^{\infty}
   e^{-\left( w(i-1)+1 \right) z} z \left( 1-e^{-z} \right)^{n-i} dz \\
 & = \frac{n!}{(i-2)! (n-i)!} \frac{w}{\sigma} (-1)^{n-i}
   \int_{0}^{\infty} z e^{-\left(w(i-1)+1\right) z} \left( e^{-z} - 1 \right)^{n-i} dz
\end{align*}
Using \cite[(3.432.1)]{grad}
\[
\int_{0}^{\infty} x^{\nu-1} e^{-mx} \left[ e^{-x} - 1 \right]^{p} dx = \Gamma(\nu) \sum_{k=0}^{p} (-1)^{k} \binom{p}{k} \frac{1}{\left( p + m - k \right)^{\nu}}
\]
with $ \nu = 2 $, $ m = w(i-1)+1 $, and $ p = n-i $, we get directly
\begin{equation} \label{appeq:fdi.gumb}
f_{D_{i,gumb}}(y)
 = \frac{n!}{(i-2)! (n-i!)} \frac{e^{y/\sigma}}{\sigma} (-1)^{n-i}
   \sum_{k=0}^{n-i} (-1)^{k} \binom{n-i}{k} \frac{1}{\left( e^{y/\sigma}(i - 1) + n - i + 1 - k \right)^{2}}
\end{equation}

\subsubsection*{Expected Spacing}
Letting $ w = y/\sigma $ the expected spacing is
\begin{align*}
E\bigl\{ D_{i,gumb} \bigr\}
 & = \int_{0}^{\infty} y f_{D_{i}}(y) dy \\
 & = \int_{0}^{\infty} \frac{n!}{(i-2)! (n-i)!} \frac{y}{\sigma} e^{y/\sigma} (-1)^{n-i} \sum_{k=0}^{n-i} (-1)^{k} \binom{n-i}{k} \frac{1}{\bigl( e^{y/\sigma} (i-1) + n-i+1-k \bigr)^{2}} dy \\
 & = \frac{n!}{(i-2)! (n-i)!} (-1)^{n-i}
  \sum_{k=0}^{n-i} (-1)^{k} \binom{n-i}{k}
  \int_{0}^{\infty} w e^{w}
    \frac{1}{\bigl( e^{w} (i-1) + n-i+1-k \bigr)^{2}} \sigma dw \\
 & = \frac{n!}{(i-2)! (n-i)!} (-1)^{n-i} \frac{\sigma}{(i-1)^{2}} \sum_{k=0}^{n-i} (-1)^{k} \binom{n-i}{k} \int_{0}^{\infty} \frac{w e^{w}}{(e^{w} + \alpha)^{2}} dw
\end{align*}
making the substitution $ \alpha = (n-i+1-k)/(i-1) $ for convenience.
The integral is done by parts, with $ u = w $, $ du = dw $, and
\begin{align*}
dv & = \frac{e^{w}}{\left( e^{w}+\alpha \right)^{2}} dw \\
v & = -\frac{1}{e^{w}+\alpha}
\end{align*}
so that
\[
\int_{0}^{\infty} \frac{w e^{w}}{\left( e^{w} + \alpha \right)^{2}} dw =
 \left[ -\frac{w}{e^{w} + \alpha}
   + \int_{0}^{\infty} \frac{dw}{e^{w} + \alpha} \right]_{0}^{\infty}
\]
The integral has the known form \cite[(2.313.1)]{grad}
\[
\int \frac{dx}{a+b e^{mx}} = \frac{1}{am} \left[ mx - \ln\left( a+b e^{mx} \right) \right]
\]
with $ a = \alpha $, $ b = 1 $, and $ m = 1 $.  Substituting back,
\begin{align*}
\int_{0}^{\infty} \frac{w e^{w}}{\left( e^{w} + \alpha \right)^{2}} dw
& = \left[ -\frac{w}{e^{w}+\alpha} + \frac{1}{\alpha} \bigl\{ w
  - \ln\left( e^{w}+\alpha \right) \bigr\} \right]_{0}^{\infty} \\
& = \left[ \frac{w e^{w}}{\alpha \left( e^{w}+\alpha \right)}
  - \frac{1}{\alpha} \ln\left( e^{w}+\alpha \right) \right]_{0}^{\infty} \\
& = \left[ \frac{w}{\alpha \left(1 + \alpha e^{-w} \right)}
  - \frac{1}{\alpha} \ln\left( e^{w}+\alpha \right) \right]_{0}^{\infty} 
\end{align*}
As $ w \rightarrow \infty $ both terms are equal and cancel, and at $ w = 0 $
only the logarithm remains, giving
\begin{equation*}
\int_{0}^{\infty} \frac{w e^{w}}{\bigl( e^{w} + \alpha \bigr)^{2}} dw = \frac{1}{\alpha} \ln (\alpha + 1) = \frac{i-1}{n-i+1-k} \ln \frac{n-k}{i-1}
\end{equation*}
Then
\begin{align} \label{appeq:edi.gumb}
E\bigl\{ D_{i,gumb} \bigr\}
 & = \frac{n!}{(i-2)! (n-i)!} (-1)^{n-i} \frac{\sigma}{(i-1)^{2}}
  \sum_{k=0}^{n-i} (-1)^{k} \binom{n-i}{k} \frac{i-1}{n-i+1-k}
   \ln\left( \frac{n-k}{i-1} \right) \nonumber \\
 & = \frac{n!}{(i-2)! (n-i)!} (-1)^{n-i} \frac{\sigma}{i-1}
  \sum_{k=0}^{n-i} (-1)^{k} \binom{n-i}{k} \frac{1}{n-i+1-k}
   \ln\left( \frac{n-k}{i-1} \right)
\end{align}
However, this form is numerically sensitive, and for $ n > 30 $ the sum
begins to diverge from numeric integration of the base equations, even
using high-precision math libraries.  More work is needed.

To simplify this to the final form, first we can re-write the factorials.
\[
\frac{n!}{(i-2)! (n-i)!} \frac{1}{i-1} = \frac{i n!}{i! (n-i)!} = i \binom{n}{i}
\]
Separating the logarithm and using $ m = n-i $ for the second series,
\begin{align*}
E\bigl\{ D_{i,gumb} \bigr\}
 = & i \binom{n}{i} (-1)^{n-i} \sigma
  \sum_{k=0}^{n-i} (-1)^{k} \binom{n-i}{k} \frac{1}{n-i+1-k} \ln(n-k) \\
 & - i \binom{n}{i} (-1)^{n-i} \sigma 
  \sum_{k=0}^{m} (-1)^{k} \binom{m}{k} \frac{1}{m-k+1} \ln(i-1)
\end{align*}
We want to put the second series in a standard form with a known value
\cite[(0.155.1)]{grad}
\[
\sum_{k=1}^{n} (-1)^{k+1} \frac{1}{k+1} \binom{n}{k} = \frac{n}{n+1}
\]
We do this by pulling out the first term, substituting $ k' = m - k $, and
adding in an extra, last term to the the sum.
\begin{align*}
\sum_{k=0}^{m} (-1)^{k} & \binom{m}{k} \frac{1}{m-k+1} \ln(i-1) \\
 & = \ln(i-1) \left\{ \frac{1}{m+1} + \sum_{k=1}^{m} (-1)^{k} \frac{m!}{k! (m-k)!} \frac{1}{m-k+1} \right\} \\
 & = \ln(i-1) \left\{ \frac{1}{m+1} + \sum_{k'=m-1}^{0} (-1)^{m-k'} \frac{m!}{(m-k')! k'!} \frac{1}{k'+1} \right\} \\
 & = \ln(i-1) \left\{ \frac{1}{m+1} + (-1)^{m} \sum_{k'=0}^{m-1} (-1)^{k'} \binom{m}{k'} \frac{1}{k'+1} \right\} \\
 & = \ln(i-1) \left\{ \frac{1}{m+1} + (-1)^{m} + (-1)^{m} \sum_{k'=1}^{m-1} (-1)^{k'} \binom{m}{k'} \frac{1}{k'+1} \right\} \\
 & = \ln(i-1) \left\{ \frac{1}{m+1} + (-1)^{m} - (-1)^{2m} \binom{m}{m} \frac{1}{m+1} + (-1)^{m} \sum_{k'=1}^{m} (-1)^{k'} \binom{m}{k'} \frac{1}{k'+1} \right\} \\
 & = \ln(i-1) \left\{ (-1)^{m} - (-1)^{m} \sum_{k'=1}^{m} (-1)^{k'+1} \binom{m}{k'} \frac{1}{k'+1} \right\} \\
 & = \ln(i-1) (-1)^{m} \left\{ 1 - \frac{m}{m+1} \right\} \\
 & = \ln(i-1) (-1)^{m} \frac{1}{m+1} \\
 & = \ln(i-1) (-1)^{n-i} \frac{1}{n-i+1}
\end{align*}
Substituting back,
\begin{equation*}
E\bigl\{ D_{i,gumb} \bigr\}
 = i \binom{n}{i} (-1)^{n-i} \sigma \left\{
  \sum_{k=0}^{n-i} (-1)^{k} \binom{n-i}{k} \frac{1}{n-i+1-k} \ln(n-k)
  -(-1)^{n-i} \frac{1}{n-i+1} \ln(i-1) \right\}
\end{equation*}
To re-write the remaining series, make the substitution $ k' = n-i-k $ so that
\begin{align*}
\sum_{k=0}^{n-i} (-1)^{k} & \binom{n-i}{k} \frac{1}{n-i+1-k} \ln(n-k) \\
 & = \sum_{k'=n-i}^{0} (-1)^{n-i-k'} \binom{n-i}{n-i-k'} \frac{1}{k'+1} \ln(i+k') \\
 & = (-1)^{n-i} \sum_{k'=0}^{n-i} (-1)^{k'} \binom{n-i}{k'} \frac{1}{k'+1} \ln(i+k')
\end{align*}
The final result is
\begin{align} \label{appeq:edi2.gumb}
E\bigl\{ D_{i,gumb} \bigr\}
 & = i \binom{n}{i} \sigma (-1)^{n-i} \left\{
  (-1)^{n-i} \sum_{k=0}^{n-i} (-1)^{k} \binom{n-i}{k} \frac{1}{1+k} \ln(i+k)
  - (-1)^{n-i} \frac{1}{n-i+1} \ln(i-1) \right\} \nonumber \\
 & = i \binom{n}{i} \sigma \left\{
  \sum_{k=0}^{n-i} (-1)^{k} \binom{n-i}{k} \frac{1}{1+k} \ln(i+k)
  -\frac{1}{n-i+1} \ln(i-1) \right\}
\end{align}


\pagebreak

\section{Matching of Logistic Expected Spacing to Quantile Estimator}
\label{app:match}

\renewcommand{\theequation}{M.\arabic{equation}}
\setcounter{equation}{0}

To show the logistic expected spacing \eqref{eq:edi.logis} matches the
estimator \eqref{eq:dq.logis}, we first want to reduce the factorials.
Place everything on a common denominator.
\begin{align*}
E\Bigl\{ D_{i,logis} \Bigr\}
 & = \frac{\sigma n!}{(i-2)! (n-i+1)!} \left\{ \left( \frac{1}{i-1} \right)^{2} - \sum_{k=1}^{n-i} \frac{(n-i-k)! (i-2)!}{(n-k)!} \right\} \\
 & = \frac{\sigma n!}{(i-2)! (n-i+1)!} \left\{ \left( \frac{1}{i-1} \right)^{2} - \sum_{k=1}^{n-i} \frac{(n-i-k)!}{\prod^{n-i+1}_{j=k} (n-j)} \right\} \\
 & = \frac{\sigma n!}{(i-2)! (n-i+1)!} \frac{1}{i-1} \left\{ \frac{1}{i-1} - \sum_{k=1}^{n-i} \frac{(n-i-k)!}{\prod^{n-i}_{j=k} (n-j)} \right\} \\
 & = \frac{\sigma n!}{(i-1)! (n-i+1)!} \frac{\prod^{n-i}_{j=1} (n-j)}{\prod^{n-i}_{j=1} (n-j)} \left\{ \frac{1}{i-1} - \sum_{k=1}^{n-i} \frac{(n-i-k)!}{\prod^{n-i}_{j=k} (n-j)} \right\} \\
 & = \frac{\sigma n!}{(n-1)! (n-i+1)!} \left\{ \frac{\prod^{n-i}_{j=1} (n-j)}{i-1} - \sum_{k=1}^{n-i} (n-i-k)! \prod^{k-1}_{j=1} (n-j) \right\} \\
 & = \frac{\sigma n}{(i-1) (n-i+1)!} \left\{ \prod^{n-i}_{j=1} (n-j) - (n-i-1)! (i-1) - \sum_{k=2}^{n-i} (i-1) (n-i-k)! \prod^{k-1}_{j=1} (n-j) \right\}
\end{align*}
In the third line we have reduced the upper product limit by factoring out
$ (i-1) $, which combines in the fourth step with $ (i-2)! $.  In the fifth
line we have combined the product introduced to the denominator with the
$ (i-1)! $ factor to get $ (n-1)! $, while the one added to the numerator
cancels the upper factors within the series, shifting the indices on its
product.  In the sixth line we have separated the $ k=1 $ term and factored
out $ i-1 $ in the denominator.  Combining the first product and the last in
the series, $ k = n-i $,
\begin{align*}
\Biggl\{ \prod_{j=1}^{n-i} (n-j) \Biggr\} & -
 \Biggl\{ (i-1) (n-i-(n-i))! \prod_{j=1}^{n-i-1} (n-j) \Biggr\} \\
 & = \big[ i - (i-1) \big] \prod_{j=1}^{n-i-1} (n-j) 
 = 1 \prod_{j=1}^{n-i-1} (n-j) 
\end{align*}
Using this result as the new first term and matching the series at
$ k = n-i-1 $,
\begin{align*}
\Biggl\{ 1 \prod_{j=1}^{n-i-1} (n-j) \Biggr\} & -
 \Biggl\{ (i-1) (n-i-(n-i-1)) \prod_{j=1}^{n-i-2} (n-j) \Biggr\} \\
 & = \big[ (i+1) - (i-1) \big] \prod_{j=1}^{n-i-2} (n-j) 
 = 2! \prod_{j=1}^{n-i-2} (n-j)
\end{align*}
Now the third step with $ k = n-i-2 $,
\begin{align*}
\Biggl\{ 2! \prod_{j=1}^{n-i-2} (n-j) \Biggr\} & -
 \Biggl\{ (i-1) (n-i-(n-i-2))! \prod_{j=1}^{n-i-3} (n-j) \Biggr\} \\
 & = 2! \big[ (i+2)-(i-1) \big] \prod_{j=1}^{n-i-3} (n-j)
 = 3! \prod_{j=1}^{n-i-3} (n-j)
\end{align*}
Continue down to $ k = 1 $.  At this point we combine our new first term and
the second term in the final $ E\Bigl\{ D_{i,logis} \Bigr\} $ equation.
\begin{align*}
\biggl\{ (n-1) (n-i-1)! \biggr\} & -
 \biggl\{ (i - 1) (n-i-1)! \biggr\} \\
 & = (n - i) (n-i-1)! = (n-i)!
\end{align*}
Multiplying by the pre-factor we get
\begin{equation} \label{appeq:dqlog.exp}
E\Bigl\{ D_{i,logis} \Bigr\}
 = \frac{\sigma n}{(i-1) (n-i+1)!} (n-i)!
 = \frac{\sigma n}{(i-1) (n-i+1)}
\end{equation}
This equation matches $ \stackrel{\sim}{E}\Bigl\{ D_{i,logis} \Bigr\} $ in
the main text.

\subsubsection*{Example of Matching}
An example might make this clearer.  Let $ i = n - 5 $.  Expanding
\eqref{appeq:edi.log} and multiplying terms to remove the denominators, we have
\begin{alignat*}{2}
E\Bigl\{ D{_i,logis} \Bigr\}
 & = \frac{\sigma n!}{6! (n-7)!} \left\{ \left( \frac{1}{n-6} \right)^{2} - \sum_{k=1}^{5} \frac{(5-k)! (n-7)!}{(n-k)!} \right\} \\
 & = \frac{\sigma n!}{6! (n-1)!} \prod^{6}_{j=1} (n-j) \left\{ \left( \frac{1}{n-6} \right)^{2} - \sum_{k=1}^{5} \frac{(5-k)! (n-7)!}{(n-k)!} \right\} \\
 & = \frac{\sigma n}{6! (n-6)} \prod_{j=1}^{6} (n-j) \left\{ \left( \frac{1}{n-6} \right) - \sum_{k=1}^{5} \frac{(5-k)! (n-7)! (n-6)}{(n-k)!} \right\} \\
 & = \frac{\sigma n}{6! (n-6)} \left\{ \prod_{j=1}^{5}(n-j) - \sum_{k=1}^{5} \frac{(5-k)! (n-1)! (n-6)}{(n-k)!} \right\} \\
 & = \frac{\sigma n}{6! (n-6)} \left\{
  \begin{aligned}
   (n-1)(n-2)(n-3)(n-4)(n-5) \\
   - 4!(n-6) \\
   - 3!(n-1)(n-6) \\
   - 2!(n-1)(n-2)(n-6) \\
   - 1!(n-1)(n-2)(n-3)(n-6) \\
   - 0!(n-1)(n-2)(n-3)(n-4)(n-6)
  \end{aligned} \right\}
\end{alignat*}
The pairs of the first and last terms, working from bottom to top, reduce
one by one to
\begin{align*}
\Bigl[ (n-5) - 0!(n-6) \Bigr] (n-1)\ldots(n-4) & = 1! (n-1)\ldots(n-4) \\
\Bigl[ 1!(n-4) - 1!(n-6) \Bigr] (n-1)\ldots(n-3) & = 2! (n-1)\ldots(n-3) \\
\Bigl[ 2!(n-3) - 2!(n-6) \Bigr] (n-1)(n-2) & = 3! (n-1) (n-2) \\
\Bigl[ 3!(n-2) - 3!(n-6) \Bigr] (n-1) & = 4! (n-1) \\
4! (n-1) - 4! (n-6) & = 5!
\end{align*}
So
\[
E\Bigl\{ D_{i,logis} \Bigr\}
 = \frac{\sigma n}{6! (n-6)} 5! = \frac{\sigma n}{6 (n-6)}
\]
which is the same as \eqref{appeq:dqlog.exp}.

\end{document}